\documentclass[11pt,fleqn]{article} 
\usepackage{latexsym}
\usepackage[T1]{fontenc}
\usepackage{amsmath}
\usepackage{amssymb}
\usepackage{bm}
\usepackage[mathcal]{euscript}
\usepackage{slashed}
\usepackage{microtype}
\usepackage{tikz}
\usetikzlibrary{arrows.meta,positioning,calc,fit}

\usepackage[top=30mm, bottom=30mm, left=25mm, right=25mm]{geometry}

\numberwithin{equation}{section}

\usepackage{indentfirst}

\newcommand\blfootnote[1]{
  \begingroup
  \renewcommand\thefootnote{}\footnote{#1}
  \addtocounter{footnote}{-1}
  \endgroup
}

\usepackage{hyperref}
\hypersetup{
	        unicode,	
	        colorlinks,
	        citecolor=blue,
	        linkcolor=blue, 
	        urlcolor=blue,
	        pdftitle={D-branes and nonlinear boundary equations in the AdS pure spinor string},
	        pdfauthor={Brenno Carlini Vallilo},
        bookmarksopen=true,
        bookmarksopenlevel=\maxdimen,
      }

\def\gl#1#2{\ifmmode \mathrm{GL}(#1; {\bf #2}) \else $\mathrm{GL}(#1; {\bf #2})$\fi}
\def\sl#1#2{\ifmmode \mathrm{SL}(#1; {\bf #2}) \else $\mathrm{SL}(#1; {\bf #2})$\fi}

\def\usp#1{\ifmmode \mathrm{USp}(#1) \else $\mathrm{USp}(#1)$\fi}
\def\spin#1{\ifmmode \mathrm{Spin}(#1) \else $\mathrm{Spin}(#1)$\fi}

\def\double #1{#1{\hbox{\kern-2pt $#1$}}}

\mathcode`\*="702A                  
\def\half{{\textstyle{1\over{\raise.1ex\hbox{$\scriptstyle{2}$}}}}}

\def \a{\alpha}
\def \b{\beta}

\def \g{\gamma}

\def \St{{\rm Str}}
\def \Tr{{\rm Tr}}

\def \Zb{{\mathbb Z}}
\def \psu{{\mathfrak{psu}}}
\def \su{\mathfrak{su}}
\def \osp{\mathfrak{osp}}
\def \so{\mathfrak{so}}
\def \spp{\mathfrak{sp}}

\def \frakg{{\mathfrak{g}}}

\def \zbar{{\bar z}}
\def \sj{{\mathsf j}}
\def \sm{{\mathsf m}}
\def \sp{{\mathsf p}}

\def \sT{{\mathsf T}}
\def \sP{{\mathsf P}}
\def \sM{{\mathsf M}}
\def \sQ{{\mathsf Q}}
\def \sK{{\mathsf K}}
\def \sD{{\mathsf D}}
\def \sR{{\mathsf R}}
\def \sS{{\mathsf S}}

\def \ST{{\mathsf{st}}}
\def \T{{\mathsf T}}
\def \im{{\rm i}}

\begin{document}

\begin{flushright}
\makebox[0pt][b]{}
\end{flushright}

\vspace{40pt}
\begin{center}
{\LARGE D-branes and nonlinear interactions 
in the $AdS$ pure spinor string}

\vspace{40pt}
Brenno Carlini Vallilo${}^{\spadesuit}$
\vspace{30pt}

${}^{\spadesuit}${\em Departamento de F\'isica y Astronom\'ia, Facultad de Ciencias Exactas,\\
  Universidad Andres Bello,
  Sazi\'e 2212, Santiago, Chile
}

\vspace{60pt}
{\bf Abstract}
\end{center}
We develop a coordinate-independent algebraic description of
zero-field boundary conditions for the pure spinor string on
$AdS_5\times S^5$ in the closed-string channel.  The construction is
organized by involutive automorphisms of
$\mathfrak{psu}(2,2|4)$ that exchange the odd sectors of the
$\mathbb Z_4$ grading.  Their fixed subalgebras provide candidate
half-BPS, Lorentzian D-brane geometries with vanishing background
world-volume flux, consistent with the standard
$AdS_2$, $AdS_3\times S^1$, $AdS_4\times S^2$, and related embeddings.
Together, the seven representatives realize the Lorentzian
zero-flux D1, D3, D5, and D7 probe sectors considered here. 
We then construct a local ansatz for the matter and ghost boundary
interactions and derive a system of consistency equations for the
tangential superconnection, transverse embedding, and four
ghost-gluing couplings.  The resulting system is a set of Dirac-Born-Infeld-like equations of motion on the brane world-volume.
\blfootnote{
${}^{\spadesuit}$ \href{mailto:vallilo@gmail.com}{vallilo@gmail.com} }

\setcounter{page}0
\thispagestyle{empty}

\newpage

\tableofcontents

\parskip = 0.1in
\section{Introduction}

The pure spinor formalism~\cite{Berkovits:2000fe}
for the superstring is an alternative to the well-known
Ramond--Neveu--Schwarz (RNS) and Green--Schwarz (GS) formalisms. Its
main advantages are manifest spacetime supersymmetry and covariant
quantization. It has passed many consistency checks and has been used
to compute string amplitudes up to
three loops~\cite{Gomez:2013sla}.
For the $AdS_5\times S^5$ background, the early development of the
formalism included the construction of vertex operators and the proof
of one-loop conformal invariance~\cite{Berkovits:2000yr,Vallilo:2002mh}.
Its integrable and BRST structures were studied through flat currents,
nonlocal charges, master symmetries, and, more recently, a derivation
from four-dimensional Chern--Simons theory
\cite{Vallilo:2003nx,Berkovits:2004jw,Chandia:2016ueo,
BerkovitsPitombo:2024}.  Quantum world-sheet analyses include 
\cite{Berkovits:2004xu,Mazzucato:2009fv,Bedoya:2010av,
Vallilo:2011fj,Ramirez:2015rma}.  Further developments of vertex
operators, harmonic superspace, and deformations can be found in
\cite{Bedoya:2010qz,Berkovits:2012ps,Chandia:2017afc,
FleuryMartins:2021}.

Despite this progress, basic questions remain about the description of
D-branes and boundary observables in the pure spinor formalism.  The
first study of the consequences of open-string boundary conditions in
this formalism was given in~\cite{Berkovits:2002ag}.
The flat-space boundary-BRST construction was subsequently extended
from the D9-brane to Abelian D$p$-branes, including the transverse
scalar superfields associated with the Dirichlet directions, in
\cite{Hanazawa:2019DBI}.  A non-Abelian extension for coincident
D$p$-branes was later obtained by introducing and quantizing boundary
fermions~\cite{Fujii:2023DBI}.  These analyses are formulated in flat
ten-dimensional superspace.  Here we instead consider half-BPS branes
in the curved $AdS_5\times S^5$ supercoset, where a
$\mathbb Z_4$-compatible involution organizes the boundary fields and
gluing conditions.
Earlier discussions of D-branes in the pure spinor formalism include Refs. \cite{Anguelova:2003sn,Schiappa:2005mk,Mukhopadhyay:2005ur}. The $AdS$ case was studied in Ref. \cite{Hanazawa:2016lvo}, building on a related analysis in the Green--Schwarz formalism \cite{Sakaguchi:2003py}. The analysis of Ref. \cite{Hanazawa:2016lvo} relies on an explicit parametrization of the coset, which obscures the coordinate-independent structure of the boundary conditions and their consequences.

Throughout this paper, ``D-brane'' means a Lorentzian half-BPS
D$p$-brane with $p\geq 1$, vanishing background world-volume gauge
field strength, and a gluing condition defined by a fixed involution.
Euclidean D-instantons and flux-deformed gluing conditions are outside
the scope of the analysis.  The boundary fields introduced below are
fluctuations about this zero-field configuration; they are not used to
add further flux-supported branches to the construction.

In the present work, we describe a coordinate-independent algebraic
framework for D-brane boundary conditions in the $AdS$ pure spinor
string.  We isolate the conditions on a gluing automorphism and relate
its fixed subalgebra to candidate half-BPS brane geometries: it has to satisfy $R\Sigma R^{-1}=\Sigma^{-1}$, where $R$ is the gluing automorphism and $\Sigma$ is the $\mathbb Z_4$ automorphism. This
distinguishes the world-sheet derivation from the classification of
half-BPS supergravity symmetry algebras in~\cite{DHoker:2008wvd}; the
latter supplies necessary symmetry data but does not, by itself,
classify boundary involutions or probe-brane embeddings.  We finally
construct the boundary interactions and derive the field-dependent
equations imposed by the boundary BRST current.

These interacting equations are the second main result of the paper.
They form a local, pure-spinor-projected Born--Infeld-type superspace
system coupling the tangential superconnection, the transverse
embedding, and the four $R$-parity-resolved ghost-gluing couplings,
while retaining the dependence on the embedding to all orders.  Their
leading terms have the
Berkovits--Pershin structure~\cite{Berkovits:2002ag}, and the higher
terms encode the curved-supercoset geometry.  To our knowledge, this is
the first local supercoset formulation of the coupled matter and ghost
boundary-BRST constraints for fluctuations about half-BPS D-branes in
the $AdS_5\times S^5$ superstring.

This paper is organized as follows. Section~\ref{sec:action} contains
an elementary review of boundary conditions in the classical bosonic
string. In Section~\ref{sec:adsboundary}, we briefly describe the pure
spinor formalism for $AdS_5\times S^5$ and discuss the conditions
imposed by the presence of world-sheet boundaries.
Section~\ref{sec:matrix-involutions} constructs the compatible
involutions and identifies their fixed brane geometries. In
Section~\ref{sec:boundinteractions}, we derive the interacting boundary
equations.
In the last section, we conclude the work and discuss possible
research directions.
Appendix~\ref{appendix:psu} contains a
review and useful descriptions of the $\psu(2,2|4)$ algebra and
also describes some of the notation used in the paper.
Appendix~\ref{appendix:osp} reviews the two orthosymplectic real forms
that occur among the retained branes and collects the diagonal-inner
and supertranspose matrix constructions used in the main text, using
the conventions of Ref.~\cite{DHoker:2008wvd}.

\section{Bosonic string in flat space}
\label{sec:action}
Before turning to boundaries in the $AdS_5\times S^5$ pure spinor
string, we isolate the same geometric problem in the flat bosonic
string using the closed-string picture.  The purpose of this example
is to show how an involution organizes the Neumann--Dirichlet split,
the preserved world-volume geometry, and the gauge and position
couplings.  We will use the same organization in the $AdS$ supercoset;
the $\mathbb Z_4$ grading, the odd currents, and the pure-spinor ghost
completion are the new ingredients there.
We will use world-sheet light-cone coordinates
$z=\tau+\sigma$ and $\bar z=\tau-\sigma$ close to the boundary. The
boundary is located at $\tau=0$. The world-sheet differential is
$d=\partial+\bar\partial=dz\partial_z+d\bar z\partial_{\bar z}$, where
$\partial_z=\frac12(\partial_\tau+\partial_\sigma)$
and $\partial_{\bar z}=\frac12(\partial_\tau-\partial_\sigma)$. The
Hodge star operator acts on the one-forms as $\star dz=dz$
and $\star d\bar z=-d\bar z$.

\subsection{Action and boundaries}

Flat 26-dimensional spacetime, with
$a=0,\ldots,25$ and
$\eta_{ab}=\operatorname{diag}(-,+,\ldots,+)$, can be
viewed as the coset $ISO(1,25)/SO(1,25)$ where $ISO(1,25)$ is
generated by $\{\sP_a,\sM_{ab}\}$ and $SO(1,25)$ is generated by
$\{\sM_{ab}\}$. The bosonic string action is given by
\begin{align}
  S = \frac{1}{\pi} \int dX^a\wedge\star d X_a.
\end{align}
We can also describe the theory using a coset element $g$ with the
parametrization $g=e^{{\rm i} X^a \sP_a}$. The global left
transformation is defined by
\begin{align}\label{globalPoincare}
  g'= e^{\im \epsilon^b\sP_b}e^{\im \omega^{ab}\sM_{ab}}e^{\im
  X^a\sP_a}= e^{\im(\epsilon^a +\Lambda^a{}_bX^b)\sP_a}e^{\im \omega^{ab}\sM_{ab}},
\end{align}
where $\Lambda^a{}_b$ is the finite Lorentz transformation generated by
$\omega^{ab}$. We have used that
\begin{align}
  e^Xe^Y=e^{\sum_{n=0}^\infty \frac{1}{n!}({\rm ad}_X)^n Y} e^X,
\end{align}
where ${\rm ad}_X Y=[X,Y]$. The right factor $e^{\im \omega^{ab}\sM_{ab}}$
in Eq.~\eqref{globalPoincare} can be eliminated by the local isotropy
transformations that define the coset
\begin{align}
  g\sim g e^{\im \phi^{ab}(z,\bar z)\sM_{ab}}.
\end{align}
Next, we define the left-invariant Maurer--Cartan current
\begin{align}
  J=g^{-1} d g = K+A=K^a\sP_a+A^{ab}\sM_{ab}= {\rm i} dX^a\sP_a,
\end{align}
which under local infinitesimal isotropy transformations
$\im \phi^{ab}(z,\bar z)\sM_{ab}$ changes to
\begin{align}
  K' + A' = K +\im [ K , \phi^{ab}(z,\bar z)\sM_{ab}] + A +
  \im [A,\phi^{ab}(z,\bar z)\sM_{ab}]+\im d\phi^{ab}\sM_{ab}.
\end{align}

The invariant coset action is
\begin{align}
  S=-\frac{1}{\pi}\int\Tr\left(K\wedge\star K\right),
\end{align}
where $\Tr(\sP_a\sP_b)=\eta_{ab}$.  To obtain the currents associated
with the global symmetries, we allow $\epsilon^a$ and $\omega^{ab}$ to
depend on the world-sheet coordinates.  The corresponding
infinitesimal variation of $K$ is
\begin{align}
  \delta K = g^{-1} (\im d\epsilon^a\sP_a +\im d\omega^{ab}\sM_{ab})g
  = \im (d\epsilon^a- d\omega^{ab}X_b)\sP_a.
\end{align}
Inserting this transformation into the action, we find that
\begin{align}
  \delta S = \frac{2}{\pi}\int \left( \star d\epsilon^a\wedge dX_a +
  \frac12 \star d\omega^{ab}\wedge (X_adX_b-X_bdX_a)\right),
\end{align}
so we identify the conserved currents
\begin{align}
  \sp^a = \frac{2}{\pi}d X^a,\quad
  \sm^{ab} =\frac{2}{\pi}\left( X^a dX^b- X^bdX^a\right),
\end{align}
which satisfy
\begin{align}
  d\star \sp^a = d\star \sm^{ab}=0,
\end{align}
when $d\star d X^a=0$. The conserved charges are
\begin{align}
  \sP^a =\oint \star \sp^a,\quad
  \sM^{ab}=\oint\star\sm^{ab}.
\end{align}

Let us consider boundary conditions and D-branes for the closed
string moving in a flat space. If we make a general variation
$\delta X^a$, the action varies as
\begin{align}
  \delta S = -\frac{2}{\pi}\int \delta X^a d\star d X_a +
  \frac{2}{\pi}\oint_{\gamma} \delta X^a\star dX_a,
\end{align}
where $\gamma$ is  the boundary at $\tau=0$.
The first term gives the equation of motion.
The boundary terms will vanish if
\begin{align}
  \delta X^a \star dX_a \Big|_{\gamma}=0
  \quad\Longleftrightarrow\quad
  \delta X^a \star\sp_a \Big|_{\gamma}=0.
\end{align}
Requiring $\delta X^a\big|_{\gamma} =0$ for all $a$ is the boundary
condition of a D(-1)-brane and requiring $\star dX_a\big|_{\gamma}=0$
for all $a$ is the boundary condition of a space-filling brane.
All other D-branes are intermediate cases of these two extremes.
A D$p$-brane is described by imposing Neumann boundary conditions on the
$\{0\cdots p\}$ directions and Dirichlet conditions on the
$\{p+1\cdots 25\}$ directions. We will denote the indices in the
directions parallel to
the D-brane by $A,B,C,\ldots$ and the directions perpendicular to the
D-brane by $A',B',C',\ldots$. We denote the corresponding coordinates
collectively by $X$ and $X'$, respectively.

We now organize this tangent--normal split algebraically.  Let
\begin{align}
 r^a{}_b=\operatorname{diag}\bigl({\bf 1}_{p+1},-{\bf 1}_{25-p}\bigr)
 \label{eq:flat-reflection}
\end{align}
act as $+1$ on the directions parallel to the brane and as $-1$ on
the transverse directions.  It extends to the Poincar\'e algebra by
\begin{align}
 R(\sP_a)=r_a{}^b\sP_b,
 \qquad
 R(\sM_{ab})=r_a{}^c r_b{}^d\sM_{cd}.
 \label{eq:flat-involution}
\end{align}
Since $r$ is a Lorentz reflection, $R$ preserves the Poincar\'e
brackets and satisfies $R^2={\bf 1}$.  Its fixed algebra is
\begin{align}
 \mathfrak{iso}(1,25)^\bullet
 =\langle\sP_A,\sM_{AB},\sM_{A'B'}\rangle
 \simeq\mathfrak{iso}(1,p)\oplus\mathfrak{so}(25-p).
\end{align}
Writing $G=ISO(1,25)$ and $H=SO(1,25)$, and defining the fixed subgroup
$G^R=\{g\in G\mid R(g)=g\}$, the corresponding orbit is
\begin{align}
 \mathcal W_{\rm flat}
 =\frac{G^R}{G^R\cap H}
 =\frac{ISO(1,p)\times SO(25-p)}
 {SO(1,p)\times SO(25-p)}
 \simeq\frac{ISO(1,p)}{SO(1,p)}.
 \label{eq:flat-brane-orbit}
\end{align}
The transverse rotation group is therefore part of the fixed symmetry,
but it belongs to the stabilizer rather than to the tangent space of
the brane.  This distinction will remain important in the $AdS$ case.

The Dirichlet boundary condition $\delta X^{A'}\big|_{\gamma}=0$
means that $X^{A'}$ is fixed along the boundary; its constant value
specifies the position of the D$p$-brane. Since the boundary lies at
fixed $\tau$, this condition is
$\partial_\sigma X^{A'}\big|_{\gamma}=0$, or equivalently
$\sp^{A'}\big|_{\gamma}=0$. We can therefore summarize all boundary
conditions in terms of the isometry currents as
\begin{align}\label{boundaryCondBos}
  \star \sp^A\big|_{\gamma}=\sp^{A'}\big|_{\gamma}=0.
\end{align}
With
\begin{align}
 P_\bullet=\frac12(1+R),
 \qquad
 P_\circ=\frac12(1-R),
\end{align}
the coset current decomposes as
\begin{align}
 K^\bullet=P_\bullet(K)=\im dX^A\sP_A,
 \qquad
 K^\circ=P_\circ(K)=\im dX^{A'}\sP_{A'}.
\end{align}
Equation~\eqref{boundaryCondBos} then takes the invariant form
\begin{align}
 \star K^\bullet\big|_\gamma=0,
 \qquad
 K^\circ\big|_\gamma=0.
 \label{eq:flat-projected-boundary-conditions}
\end{align}
The restriction to $\gamma$ includes the pullback.  Since the boundary
is at fixed $\tau$, these equations set $K_\tau^\bullet=0$ and
$K_\sigma^\circ=0$.  Thus the fixed and anti-fixed currents encode the
Neumann and Dirichlet conditions, respectively, in a form that is
independent of the coset representative.

There is another way of defining the D-brane without restricting the
variations at the boundary. We add a boundary action
\begin{align}\label{boundaryInt1}
  S_b= -\frac{2}{\pi}\oint_{\gamma} (X^{A'}-x_0^{A'}) \mathsf{T
  }_{A'},
\end{align}
where $x_0^{A'}$ is a constant vector in the transverse directions and
$\mathsf{T}_{A'}$ is an auxiliary boundary one-form. If we now vary the
action together with the boundary interaction, the remaining terms are
\begin{align}
  \delta(S+S_b)=& -\frac{2}{\pi}\int \delta X^a d\star d X_a +
  \frac{2}{\pi}\oint_{\gamma} \delta X^A\star dX_A +\frac{2}{\pi}\oint_{\gamma}\delta X^{A'}\left( \star d X_{A'} - \mathsf{T}_{A'}\right)\\ &-
  \frac{2}{\pi}\oint_{\gamma}(X^{A'}-x_0^{A'})\delta \mathsf{T}_{A'}.
\end{align}
Therefore, stationarity of
$S+S_b$ for arbitrary $\delta X^A$, $\delta X^{A'}$ and $
\delta \mathsf{T}_{A'}$ gives the bulk
equations of motion and the
following conditions at $\gamma$
\begin{align}
  \star dX^A\big|_\gamma=0,\quad
  \mathsf{T}_{A'}=\star dX_{A'}\big|_\gamma,\quad
  X^{A'}\big|_\gamma=x^{A'}_0.
\end{align}
These equations describe a D$p$-brane located at $x^{A'}_0$ and fix
the auxiliary boundary one-form. Note that the last boundary condition
implies that $K^\circ\big|_\gamma=0$. The
boundary interaction breaks Poincar\'e invariance in the transverse
directions. We now show how this invariance can be restored by
allowing the D-brane to be a dynamical object.

\subsection{Boundary vertex operators}

D-branes carry collective degrees of
freedom~\cite{Li:1995pq,Callan:1995xx}. We consider the
simple case in which all boundary components $\gamma$ are attached to
D-branes of the same dimension. More generally, each boundary
component may carry an independent set of boundary conditions. The
collective degrees of freedom of the branes are described by a boundary
interaction of the form
\begin{align}\label{boundaryI}
  I_{b}=\frac{2}{\pi}\oint_{\gamma}\Big(
  \bm{A}_A(X)d X^A +
  \left(\bm{\Phi}^{A'}(X)-X^{A'}\right)\mathsf{T}_{A'}\Big),
\end{align}
where $\bm{A}_A$ and $\bm{\Phi}^{A'}$ are functions only of $X^A$, the
coordinates of the world-volume. The field
$\bm{A}_A$ describes gauge fields living on the brane,
and $\bm{\Phi}^{A'}$ are the transverse embedding fields
of the brane. This interaction generalizes
Eq.~\eqref{boundaryInt1} and is invariant under the gauge transformations
\begin{align}
  \delta \bm{A}_A = \partial_A \Lambda(X),\quad
  \delta \bm{\Phi}^{A'} = 0,
\end{align}
where $\Lambda(X)$ is a function of the world-volume coordinates.

We can now discuss how the presence of Eq.~\eqref{boundaryI} changes the
boundary conditions. If we compute the variation of $S+I_b$ we get
\begin{align}
  \delta(S+I_b)=&-\frac{2}{\pi}\int \delta X^a d\star d X_a\\
  &+ \frac{2}{\pi}\oint_{\gamma} \delta X^A\big(\star dX_A
  -dX^B\bm{F}_{BA} +\mathsf{T}_{C'}\partial_A\bm{\Phi}^{C'}\big)\\
  &+\frac{2}{\pi}\oint_{\gamma}\left( \delta X^{A'}\left(\star d X_{A'}-\mathsf{T}_{A'}\right)+\left(\bm{\Phi}^{A'}-X^{A'}\right) \delta\mathsf{T}_{A'}\right).
\end{align}
Here $\bm{F}_{AB}=\partial_A\bm{A}_B-\partial_B\bm{A}_A$. The
boundary terms will vanish for arbitrary $\delta X^A$, $\delta X^{A'}$,
and $\delta\mathsf{T}_{A'}$ if
\begin{align}
  &\big(\star dX_A-dX^B\bm{F}_{BA} +
    \mathsf{T}_{C'}\partial_A\bm{\Phi}^{C'}\big)\big|_\gamma=0,\\
  &X^{A'}\big|_\gamma=\bm{\Phi}^{A'}, \quad \mathsf{T}_{A'}\big|_\gamma=\star dX_{A'}\big|_\gamma.
\end{align}
Thus $\bm{\Phi}^{A'}$ specifies the embedding of the
D-brane in the transverse directions. The Poincar\'e
invariance in the transverse directions can be restored by declaring
that $\bm{\Phi}^{A'}$ transforms inhomogeneously under translations in
the transverse directions
\begin{align}
  \delta \bm{\Phi}^{A'} = \epsilon^{A'},
\end{align}
where $\epsilon^{A'}$ is the translation parameter. This inhomogeneous transformation is closely related to the nonlinearly realized supersymmetries characteristic of Born--Infeld equations of motion \cite{Berkovits:2002ag}.

Quantum conformal invariance of the full theory, including
the boundary vertex operator in Eq.~\eqref{boundaryI}, determines the
dynamics of the fields $\bm{A}_A(X)$ and $\bm{\Phi}^{A'}(X)$
\cite{Li:1995pq,Callan:1995xx}. In the pure spinor formalism,
conservation of the boundary BRST charge gives a supersymmetric analogue
of the conditions obtained from the bosonic sigma-model beta function.
In flat space this mechanism yields the Born--Infeld dynamics of the
brane gauge multiplet using classical boundary BRST
invariance~\cite{Berkovits:2002ag}.

\section{\texorpdfstring{$AdS$}{AdS} pure spinor string with boundaries}
\label{sec:adsboundary}

We now apply the same language to the $AdS$ pure spinor string. We
mostly use the notation and conventions of
\cite{Chandia:2016ueo,Ramirez:2015rca}. The pure spinor action on
$AdS_5\times S^5$ is~\cite{Berkovits:2000fe,Berkovits:2000yr,Vallilo:2002mh}
\begin{align}\label{action}
  S=-\frac{1}{\pi} \int\! \St\left( K_2\wedge\star K_2 +
  K_1\wedge K_3 +4 N\wedge \bar N -4 \bar\nabla \lambda\wedge\omega -
 4\nabla\bar\lambda\wedge\bar\omega  \right).
\end{align}
Let us explain the notation. The one-form currents
$K_i$ are constructed using an element $g$ of the coset
$G/H_0$, where
$G=PSU(2,2|4)$ and $H_0=USp(2,2)\times USp(4)$:
\begin{align}
  g^{-1}dg = J=A+K= A +K_1+K_2+K_3,
\end{align}
where we have used the $\Zb_4$ decomposition
$\frakg=\frakg_0\oplus\frakg_1\oplus\frakg_2\oplus\frakg_3$ of
$\psu(2,2|4)$. We write the full one-forms as
\begin{align}
 K_i=K_{i,z}\,dz+K_{i,\bar z}\,d\bar z.
\end{align}
Thus $K_i$ without a component label denotes the full one-form, while
$K_{i,z}$ and $K_{i,\bar z}$ denote its $dz$ and $d\bar z$
coefficients.
The same convention applies to the grade-zero connection: in
$\nabla$ and $\bar\nabla$, $A$ and $\bar A$ denote its $dz$ and
$d\bar z$ coefficients, respectively.
The ghosts
$\lambda$ and $\bar\lambda$ have $\mathbb{Z}_4$ charges $1$ and $3$,
respectively, and satisfy the pure spinor conditions
\begin{align}
  \{\lambda,\lambda\}=\{\bar\lambda,\bar\lambda\}=0.
\end{align}
Their conjugate momenta $\omega$ and $\bar\omega$ are holomorphic and
antiholomorphic one-forms of $\mathbb{Z}_4$ charges $3$ and $1$ and have the gauge transformations
\begin{align}
  \delta\omega = [X,\lambda],\quad
  \delta\bar\omega = [\bar X,\bar\lambda],  
\end{align} 
where $X$ and $\bar X$ $\in \mathfrak{g}_2$. The
ghost one-form currents are $N=-\{\lambda,\omega\}$ and
$\bar N=-\{\bar\lambda,\bar\omega\}$. When their form factors are
displayed explicitly, we use the same symbols for their nonzero chiral
coefficients. Accordingly, the BRST transformations below are written
for these coefficients, whereas expressions with wedge products use
the full one-forms. Finally, on an adjoint-valued field $X$,
\begin{align}
 \nabla X=\partial X+[A,X],
 \qquad
 \bar\nabla X=\bar\partial X+[\bar A,X].
\end{align}

The BRST-like transformations are
\begin{align}
  Q g =
  g(\lambda+\bar\lambda),\quad Q\lambda=Q\bar\lambda=0,\quad
  Q\omega =-K_{3,z},\quad Q\bar\omega = -K_{1,\bar z}.
\end{align}
Consequently the Lorentz currents transform symmetrically,
\begin{align}
  QN=[K_{3,z},\lambda],
  \qquad
  Q\bar N=[K_{1,\bar z},\bar\lambda].
  \label{eq:brst-lorentz-currents}
\end{align}
The action, ghost equations, and global current below use this
symmetric definition of the two Lorentz currents.
In this representative, the BRST transformation is nilpotent only
modulo a local grade-zero transformation, the equations of motion, and
the pure-spinor gauge symmetry. In particular,
$Q^2g=g\Lambda_0$, with
$\Lambda_0=\{\lambda,\bar\lambda\}\in\frakg_0$,
$Q^2\omega=-\nabla\bar\lambda+[\lambda,K_{2,z}]$, and
$Q^2\bar\omega=-\bar\nabla\lambda+[\bar\lambda,K_{2,\bar z}]$.
Thus $Q^2g$ is precisely a right local $H_0$ transformation of the
coset representative.  On the conjugate ghosts, closure additionally
uses their equations of motion and the pure-spinor gauge equivalence.
Equivalent off-shell descriptions are discussed in
\cite{Berkovits:2007rj,Bedoya:2010qz,Chandia:2014sta}. Below, BRST
statements are understood on shell and on local-$H_0$- and
pure-spinor-gauge-invariant quantities, on which $Q$ is nilpotent.

In the closed-string channel, D-branes are localized objects that can emit or
absorb closed strings. The boundaries of the world-sheet in this
picture are space-like and represent the moment of creation or
annihilation of the string. A channel rotation relates the fixed-$\tau$
conventions used here to the timelike open-string boundary used in
\cite{Berkovits:2002ag}; all gluing signs below follow the closed-channel
orientation. Under general
variations $\delta g =\Omega g$, $\delta\lambda$,
$\delta\omega$, $\delta\bar\lambda$, $\delta\bar\omega$  the action
varies as
\begin{align}\label{boundaryS}
  \delta S =&\; \frac{2}{\pi} \int\St\Big( (d\star{\sj})\Omega -2(\bar\nabla\lambda-[\bar N,\lambda])\wedge\delta\omega -2
  (\nabla\bar\lambda-[ N,\bar\lambda])\wedge\delta\bar\omega\\
  &+2(\bar\nabla\omega-[\bar N,\omega])\delta\lambda
  +2(\nabla\bar\omega-[ N,\bar\omega])\delta\bar\lambda \Big)
  \\&+\frac{2}{\pi}\oint_{\gamma}\St\big(\star{\sj}\Omega+
  2\omega\delta\lambda+2\bar\omega\delta\bar\lambda \big),
 \end{align}
 where the first two lines give the equations of motion and the last
 line is the boundary term. The one-form current
\begin{align}\label{globalC}
  \sj ={}& g\left( K_{1,z} + 2K_{2,z} + 3 K_{3,z} +4N \right)g^{-1}\,dz
  \nonumber\\
  &+g\left(3 K_{1,\bar z}+2K_{2,\bar z} + K_{3,\bar z}
  +4\bar N \right)g^{-1}\,d\bar z,
\end{align}
is the conserved current associated with the global
$PSU(2,2|4)$ symmetry of the sigma model, and its conservation
$d\star\sj=0$ gives the equations of motion for the coset element $g$. The other
terms are the equations of motion for the ghosts. For later
convenience, we define a one form $L$ as
\begin{align}\label{defL}
  L=L_z dz + L_{\bar z}d\bar z= (K_{1,z} + 2K_{2,z} + 3 K_{3,z})dz +
  (3 K_{1,\bar z}+2K_{2,\bar z} + K_{3,\bar z})d\bar z.
\end{align}

The conserved current obeys the BRST transformation
\cite{Berkovits:2004jw}
\begin{align}\label{brstj}
  Q \sj = \star d \big( g(\lambda -\bar\lambda)g^{-1}\big) +
  4g\big(\nabla\bar\lambda -[N,\bar\lambda] \big)g^{-1}dz +
  4g\big(\bar\nabla\lambda-[\bar N,\lambda] \big)g^{-1}d\bar z.
\end{align}
This relation ensures the on-shell BRST invariance of the global charges
\begin{align}
\sQ_\psu =\oint  \star \sj.
\end{align}

The first two lines of Eq.~\eqref{boundaryS} give the equations of motion
while the last gives the boundary conditions.
Considering first the matter part, at the boundary we require
\begin{align}\label{boundTerm}
  \oint_{\gamma}\St\big(\star{\sj}\,\Omega\big)=0.
\end{align}
First take $\Omega$ to be arbitrary at the boundary. Since the boundary lies at fixed $\tau$, $dz=d\sigma$ and $d\bar z=-d\sigma$ there. Writing in components we see that the current has to satisfy
\begin{align}\label{eq:gluing}
  \sj_z\Big|_{\gamma}=-\sj_\zbar\Big|_{\gamma}.
\end{align}

This resembles the ``gluing condition'' of WZW
models~\cite{Alekseev:1998mc,Stanciu:1999nx}, but
here it is a Neumann-type boundary condition, since the variation of $g$ is
allowed to be free in all directions. This corresponds to the
formal all-Neumann gluing of a space-filling brane.  It is useful to
distinguish this variational statement from supersymmetry: the explicit
zero-field fermionic gluing constraints of~\cite{Hanazawa:2016lvo}
exclude $p=9$ in the Lorentzian ansatz used here.  In their conventions,
the fermionic gluing condition requires an even number of Neumann directions
inside $AdS_5$ for $p=1\pmod 4$, whereas a D9 has five.  The absence of a
sixteen-supercharge fixed algebra containing the full
$\su(2,2)\oplus\su(4)$ bosonic isometry is the corresponding symmetry
cross-check.  We therefore retain Eq.~\eqref{eq:gluing} only as a formal
all-Neumann boundary condition, not as a half-BPS D9-brane.  This does
not make a claim about non-supersymmetric or field-dependent gluing
conditions.

In the closed-string channel, conservation of the BRST charge requires
the left- and right-moving BRST fluxes to agree at the boundary.  In
our conventions this condition is\footnote{See
\cite{Berkovits:2002ag} for the analogous result in the open-string
channel.}
\begin{align}\label{eq:brst-current-matching}
 \mathcal B_{\rm PS}
 :=\St\left(\lambda K_{3,z}+\bar\lambda K_{1,\bar z}\right)
 \Big|_\gamma=0.
\end{align}
This unreduced form must be retained until the interacting matter
boundary equations have been imposed.  For the field-independent
problem considered below, the free momentum equation and the fixed
ghost gluing reduce Eq.~\eqref{eq:brst-current-matching} to
\begin{align}\label{eq:brstglue}
  \St\Big( (\lambda-\bar\lambda)g^{-1}\partial_\sigma g \Big)\Big|_{\gamma}=0.
\end{align}
We derive the reduction and its interacting replacement in
Section~\ref{sec:boundinteractions}.

The zero-field conditions in Eqs. \eqref{boundTerm} and \eqref{eq:brstglue} are invariant under the local isotropy symmetry; this motivates the left variation of the coset element used above. We now determine when both conditions admit a solution. Reference \cite{Hanazawa:2016lvo} studied only Eq. \eqref{eq:brstglue} using an explicit coset parametrization. Here we treat both conditions in a coordinate-independent form.

\subsection{\texorpdfstring{$D$-branes}{D-branes} from the boundary conditions}

For arbitrary $\Omega$, the gluing condition describes a formal
space-filling boundary state. We now ask which lower-dimensional,
half-BPS boundary conditions can be described by a restricted set of
variations. Specifically, for different choices of $\Omega$, $\delta\lambda$, and $\delta\bar\lambda$, we impose the following two conditions:
\begin{align}\label{generalB}
  \St\big(\star{\sj}\Omega+
  2\omega\delta\lambda+2\bar\omega\delta\bar\lambda \big)\Big|_{\gamma}=0,\quad
\St\Big( (\lambda-\bar\lambda)g^{-1}\partial_\sigma g \Big)\Big|_{\gamma}=0.
\end{align}

The ghost part of the first supertrace can be solved by imposing
\begin{align}\label{ghostBC}
  \delta\bar\lambda\Big|_{\gamma} = R(\delta\lambda)\Big|_{\gamma},
  \quad
  \bar\omega \Big|_{\gamma} = -R(\omega)\Big|_{\gamma}.
\end{align}
Here $R$ is an automorphism of $\psu(2,2|4)$. The ghost boundary term
vanishes when $R$ is an involution that preserves the Lie bracket,
supertrace, real form, and local grade-zero algebra, and exchanges
$\frakg_1$ with $\frakg_3$.\footnote{Using the $10d$ description of
the algebra briefly reviewed in Appendix~\ref{appendix:psu}, a
representation for this involution is
$R(\delta\lambda) = R(\delta\lambda^\alpha\sQ_\alpha) =
  \delta\lambda^\alpha R(\sQ_\alpha)=
  \delta\lambda^\alpha R_\alpha{}^\beta \bar\sQ_\beta$.
Reference~\cite{Hanazawa:2016lvo} derives the constraints on
$R_\alpha{}^\beta$ in the ten-dimensional description and gives
explicit representatives.} For clarity, the discussion below treats
complex-linear involutions. Antilinear maps require a separate
real-linear decomposition and are outside the scope of this paper.
These conditions define candidate zero-flux half-BPS boundary
conditions.  In Section~\ref{sec:matrix-involutions} we construct one
compatible representative for every retained probe sector.
 Before analyzing this condition, let us continue with the
discussion for a general involution. As usual, the condition on the
variation of the ghosts will be interpreted as coming from a condition
on the ghosts themselves,
\begin{align}\label{ghostgluing1}
  \bar\lambda\Big|_{\gamma} = R(\lambda)\Big|_{\gamma}.
\end{align}
The symmetric Lorentz-current definitions then imply
\begin{align}
  \bar N\Big|_\gamma=-R(N)\Big|_\gamma.
  \label{eq:lorentz-current-gluing}
\end{align}
As in the case of the bosonic string, we will later discuss how to keep the
variations of the fields independent. Since $R$ is an involution, we
can define its positive- and negative-eigenvalue subspaces,
\begin{align}
  X^\circ=P_\circ(X) =\frac12( X - R(X)),\quad
  X^\bullet=P_\bullet(X) =\frac12( X + R(X)),
\end{align}
and decompose $\frakg=\frakg^\bullet\oplus\frakg^\circ$. This is a
$\mathbb Z_2$ grading, with
\begin{align}
  [\frakg^\bullet,\frakg^\bullet]&\subseteq\frakg^\bullet, &
  [\frakg^\circ,\frakg^\circ]&\subseteq\frakg^\bullet, &
  [\frakg^\bullet,\frakg^\circ]&\subseteq\frakg^\circ,
\end{align}
and the supertrace satisfies
\begin{align}
  \St(\frakg^\bullet\frakg^\circ)=0.
\end{align}
For the even grades, $R$ preserves $\frakg_0$ and $\frakg_2$, so each
splits into fixed and anti-fixed parts. For the odd grades, $R$ pairs
$\frakg_1$ with $\frakg_3$; a symbol such as
$\frakg_1^\bullet$ is shorthand for the diagonal combination
$\frac12(\Psi+R(\Psi))$, with $\Psi\in\frakg_1$ and
$R(\Psi)\in\frakg_3$, rather than a subspace of $\frakg_1$ alone.
We assume that $R$ integrates to an involutive automorphism of the
identity component of $G$, denoted by the same letter, and define
\begin{align}
 G^R=\{g\in G\mid R(g)=g\}.
 \label{eq:fixed-subgroup-definition}
\end{align}

The condition on the ghosts can now be expressed as
\begin{align}\label{ghostBCnew}
  \bar\lambda^\circ=-\lambda^\circ,\quad
  \bar\lambda^\bullet=\lambda^\bullet,\quad
  \bar\omega^\circ=\omega^\circ,\quad \bar\omega^\bullet=-\omega^\bullet.
\end{align}

We now return to the matter boundary term and decompose the variation of the coset element as
\begin{align}
  \Omega = \Omega^\bullet +\Omega^\circ.
\end{align}
The matter boundary condition becomes
\begin{align}
  \St\big(\star{\sj^\circ}\Omega^\circ+
  \star{\sj^\bullet}\Omega^\bullet\big)\Big|_\gamma=0.
\end{align}
The fixed subspace $\frakg^\bullet$ is the preserved symmetry
superalgebra. Its grade-two component gives physical Neumann directions,
whereas $\frakg_0^\bullet$ belongs to the stabilizer. Similarly,
$\frakg_2^\circ$ gives physical Dirichlet directions, while
$\frakg_0^\circ$ describes broken rotations of the embedding.  Let
$H_0$ be the grade-zero isotropy group.  For the reference brane through
the identity coset,
\begin{align}\label{eq:tangent-normal-split}
  \mathcal W_0&=G^R/(G^R\cap H_0), &
  T_{[e]}\mathcal W_0&\simeq\frakg_2^\bullet, &
  N_{[e]}\mathcal W_0&\simeq\frakg_2^\circ.
\end{align}
In particular, a D$p$-brane representative must satisfy
$\dim\frakg_2^\bullet=p+1$.  The full quotient $G/G^R$ is not the normal
space: it also contains the broken grade-zero rotations.

With this distinction understood, the matter boundary term vanishes if
\begin{align}
  \star{\sj^\bullet}\Big|_\gamma=0, \quad \Omega^\circ\Big|_\gamma=0.
\end{align}
Locally, if the group element at the boundary lies in the identity
component of the fixed subgroup and has the form
$g=e^{X^\bullet(\sigma)}$, this condition is equivalent to
\begin{align}
  \star{L^\bullet}\Big|_\gamma=0,
\end{align}
because the ghost contribution inside
$(\star\sj)\big|_\gamma$ is proportional to
$N+\bar N$.  Equation~\eqref{eq:lorentz-current-gluing} gives
\begin{align}
  P_\bullet(N+\bar N)
  =P_\bullet\bigl(N-R(N)\bigr)=0.
  \label{eq:fixed-ghost-current-cancellation}
\end{align}
Thus the fixed ghost contribution vanishes with the same convention
used in the action and ghost equations. Later we will allow for a more
general boundary condition.

We first carry out the argument for the reference brane
$\mathcal W_0$.  The coset element evaluated at the boundary then has
the local form
\begin{align}
  g(\tau,\sigma)\Big|_\gamma=e^{X^\bullet(\sigma)},
\end{align}
where $X^\bullet(\sigma)\in \frakg^\bullet$. From this we see that
$g^{-1}\partial_\sigma g\in \frakg^\bullet$, as expected.  If we use
Eq.~\eqref{ghostBCnew} in the second condition of
Eq.~\eqref{generalB}, we have
\begin{align}
  2\St\Big(\lambda^\circ g^{-1}\partial_\sigma g \Big)\Big|_{\gamma}=0,
\end{align}
since $\St(\frakg^\circ\frakg^\bullet)=0$.
An equivalent way to see this is that the condition
$\star\sj^\bullet\big|_{\gamma}=0$ is compatible with the BRST
transformation in Eq.~\eqref{brstj}, since on shell
\begin{align}
  Q\star \sj\Big|_{\gamma}=2 \partial_\sigma
  \big( g(\lambda^\circ) g^{-1})\Big|_{\gamma}
  \in \frakg^\circ.
\end{align}

A translated brane is described without mixing left and right orbit
conventions.  If $g_0\in G$, define
\begin{align}
  \widehat g=g_0^{-1}g,\qquad
  R_{g_0}=\operatorname{Ad}_{g_0}\,R\,\operatorname{Ad}_{g_0}^{-1},
  \qquad G^{R_{g_0}}=g_0G^Rg_0^{-1}.
\end{align}
The boundary condition is $\widehat g|_\gamma\in G^R$ in a local coset
section, while global-frame quantities are projected with $R_{g_0}$.
The translated orbit is
\begin{align}
  \mathcal W_{g_0}
  =G^{R_{g_0}}/(G^{R_{g_0}}\cap g_0H_0g_0^{-1})
  \simeq G^R/(G^R\cap H_0),
\end{align}
and its tangent and normal spaces are the left translates of those in
Eq.~\eqref{eq:tangent-normal-split}.

\subsection{Boundary action with independent boundary fields}
The restricted variations $\Omega^\circ|_\gamma=0$ together with
Eq.~\eqref{ghostBCnew} already give a well-posed variational principle.
As in the bosonic warm-up, we can instead keep all boundary variations
independent by introducing an auxiliary boundary one-form.  For this
purpose it is more natural to return to the local frame of the
Maurer--Cartan current.  We write
\begin{align}
 \begin{aligned}
 \delta g&=g\Theta,
 &
 \Theta&:=g^{-1}\delta g=A_\delta+\Omega',
 \\
 \Omega'&:=P_{\mathfrak m}\Theta\in\mathfrak m,
 &
 P_{\mathfrak m}&:=1-P_0,
 \\
 \mathfrak m&=\frakg_1\oplus\frakg_2\oplus\frakg_3.
 \end{aligned}
 \label{eq:right-coset-variation}
\end{align}
Here $A_\delta=P_0\Theta$ is the grade-zero part of the full body-fixed
variation.  It is needed when differentiating a chosen coset section,
although it does not pair with the coset momentum in the boundary term.
We use
\begin{align}
 g^{-1}dg&=A+K,
 &
 g^{-1}\sj g
 &=L+4N\,dz+4\bar N\,d\bar z,
 \label{eq:right-frame-current-identities}
\end{align}
where $A\in\frakg_0$ is the isotropy connection,
$K\in\mathfrak m$ is the Maurer--Cartan coset current, and
$D_A=d+[A,\,\cdot\,]$. On the ghost equations of motion,
\begin{align}
 \bar\nabla N=[\bar N,N],
 \qquad
 \nabla\bar N=[N,\bar N],
\end{align}
the covariant derivative of
$\star(N\,dz+\bar N\,d\bar z)$ vanishes. The right-frame equation of
motion therefore becomes
\begin{align}
 D_A\star L+[K,\star L]
 +4[K,\star(N\,dz+\bar N\,d\bar z)]
 =0.
 \label{eq:right-frame-bulk-equation}
\end{align}
At the boundary, the grade-zero Lorentz currents do not pair with
$\Omega'\in\mathfrak m$. We therefore define the
right-frame canonical momentum directly by
\begin{align}
 \mathcal P
 :=\left.\star L\right|_\gamma.
 \label{eq:right-frame-boundary-momentum}
\end{align}
This is the momentum conjugate to a right variation of $g$; it is not
a second Maurer--Cartan velocity.

The local-frame Cartan element is
\begin{align}
 \mathcal C_R(g)=g^{-1}R(g).
 \label{eq:local-Cartan-element-draft}
\end{align}
This is the usual Cartan map for the left quotient $G^R\backslash G$:
it is invariant under $g\mapsto kg$ with $k\in G^R$. Near the reference
orbit, we choose a local $H_0$ gauge in which
\begin{align}
 g=g_\bullet e^{Y^\circ},
 \qquad
 g_\bullet\in G^R,
 \qquad
 Y^\circ\in\mathfrak m^\circ
 :=\mathfrak m\cap\frakg^\circ,
 \qquad
 R(Y^\circ)=-Y^\circ.
 \label{eq:local-fixed-normal-factorization}
\end{align}
The restriction to $\mathfrak m^\circ$ is part of the local $H_0$
gauge choice: it removes the broken grade-zero rotations and leaves
only the physical anti-fixed coset directions.
Then $\mathcal C_R(g)=e^{-2Y^\circ}$, so its logarithm gives the
transverse coordinate directly in the Maurer--Cartan frame. The
remaining isotropy transformations are $h\in H_0\cap G^R$ and act by
\begin{align}
 g_\bullet\longmapsto g_\bullet h,
 \qquad
 Y^\circ\longmapsto\operatorname{Ad}_{h^{-1}}Y^\circ.
 \label{eq:residual-isotropy-factorization}
\end{align}
Choose an $R$-invariant symmetric neighbourhood $U$ of $0$ in $\mathfrak g$ on which the exponential map is one-to-one, and denote the inverse of $\exp|_U:U\to\exp(U)$ by $\operatorname{Log}_U$.  For a coset representative $g$ such that
$\mathcal C_R(g)\in\exp(U)$, define
\begin{align}
 Y^\circ(g)
 =-\frac12\operatorname{Log}_U\mathcal C_R(g).
 \label{eq:local-Cartan-position}
\end{align}
This definition is local: it applies only while
$\mathcal C_R(g)$ remains in the chosen logarithmic branch.  Since
$R(\mathcal C_R)=\mathcal C_R^{-1}$, it follows that
$R(Y^\circ)=-Y^\circ$.  Under an isotropy transformation
$g\mapsto gh$, with $h\in H_0\cap G^R$, it transforms as
\begin{align}
 \mathcal C_R(gh)
 =h^{-1}\mathcal C_R(g)h,
 \qquad
 Y^\circ(gh)=\operatorname{Ad}_{h^{-1}}Y^\circ(g).
\end{align}
Thus $Y^\circ$ is a local-frame variable covariant under the residual
isotropy group of the brane.

The variation of this coordinate is also exact.  The body-fixed
variation $\Theta=g^{-1}\delta g$ is obtained by multiplying on the
left, whereas the Cartan element naturally appears below in
the combination $\delta\mathcal C_R\,\mathcal C_R^{-1}$.  We therefore
use the differential of the exponential map translated back to the
identity from the right:
\begin{align}
 \delta(e^X)e^{-X}
 &=\operatorname{dexp}_X(\delta X),
 &
 \operatorname{dexp}_X
 &:=\int_0^1ds\,e^{s\operatorname{ad}_X}
 =\frac{e^{\operatorname{ad}_X}-1}{\operatorname{ad}_X}.
 \label{eq:right-trivialized-dexp}
\end{align}
The quotient denotes its power series and is regular at
$\operatorname{ad}_X=0$.  Thus ``right-trivialized'' refers only to the
factor $e^{-X}$ on the right of the varied exponential.
Multiplication on the left would instead give
$e^{-X}\delta(e^X)=\operatorname{dexp}_{-X}(\delta X)$; keeping the side
explicit avoids any ambiguity between the two conventions.
The inverse of \(\operatorname{dexp}_X\) exists provided
\begin{align}
 \operatorname{spec}(\operatorname{ad}_X)\cap
 2\pi\im\bigl(\mathbb Z\setminus\{0\}\bigr)=\varnothing.
 \label{eq:dexp-invertibility}
\end{align}
In particular, $\operatorname{ad}_X$ must not have an eigenvalue equal
to a nonzero integer multiple of $2\pi\im$.  Varying the Cartan element
then gives
\begin{align}
 \delta\mathcal C_R
 &=-\Theta\mathcal C_R+\mathcal C_RR(\Theta),
 &
 \delta\mathcal C_R\,\mathcal C_R^{-1}
 &=-\Theta+\operatorname{Ad}_{\mathcal C_R}R(\Theta).
 \label{eq:right-Cartan-element-variation}
\end{align}
Since $\mathcal C_R=e^{-2Y^\circ}$, the definition in
Eq.~\eqref{eq:right-trivialized-dexp} gives the chain rule in the form
\begin{align*}
 \delta\mathcal C_R\,\mathcal C_R^{-1}
 =\operatorname{dexp}_{-2Y^\circ}(-2\delta Y^\circ).
\end{align*}
Comparing this result with
Eq.~\eqref{eq:right-Cartan-element-variation} and applying the inverse
differential gives
\begin{align}
 \delta Y^\circ
 =\frac12
 \left(\operatorname{dexp}_{-2Y^\circ}\right)^{-1}
 \left(
 \Theta-\operatorname{Ad}_{\mathcal C_R}R(\Theta)
 \right).
 \label{eq:right-position-variation}
\end{align}
The apparent singularity at $Y^\circ=0$ is removable.  On the reference
brane Eq.~\eqref{eq:right-position-variation} reduces to
\begin{align}
 \left.\delta Y^\circ\right|_{\mathcal W_0}
 =P_\circ\Theta=\Omega'^\circ,
 \label{eq:right-linear-position-variation}
\end{align}
for variations within the local factorization
\eqref{eq:local-fixed-normal-factorization}.  The grade-zero part is
then fixed by the section, while the independent anti-fixed variation
lies in $\mathfrak m^\circ$.  This is the expected transverse
variation.

The boundary contribution from the variation of the bulk action is
\begin{align}
 \left.\delta S\right|_\gamma
 =\frac{2}{\pi}\oint_\gamma
 \St\left(
 \mathcal P^\bullet\Omega'^\bullet
 +\mathcal P^\circ\Omega'^\circ
 +2\omega\delta\lambda
 +2\bar\omega\delta\bar\lambda
 \right).
 \label{eq:right-bulk-boundary-variation-split}
\end{align}
Let $\mathsf T^\circ$ be an auxiliary boundary one-form in the same
local anti-fixed coset bundle as $Y^\circ$.  In particular,
$\mathsf T^\circ\mapsto\operatorname{Ad}_{h^{-1}}\mathsf T^\circ$
under $h\in H_0\cap G^R$, just as $Y^\circ$ and
$\mathcal P^\circ$ do. Invariance of the supertrace then gives
\begin{align}
 \St\left(
 \operatorname{Ad}_{h^{-1}}Y^\circ\,
 \operatorname{Ad}_{h^{-1}}\mathsf T^\circ\right)
 =\St\left(Y^\circ\mathsf T^\circ\right).
\end{align}
Thus the position coupling is invariant under the residual isotropy
transformations for which $R(h)=h$. The AdS analogue of
Eq.~\eqref{boundaryInt1} is the first-order position term
\begin{align}
 I_{\rm pos}
 =-\frac{2}{\pi}\oint_\gamma
 \St\left(Y^\circ\mathsf T^\circ\right).
 \label{eq:AdS-first-order-position}
\end{align}
Its ordinary variation is
\begin{align}
 \delta I_{\rm pos}
 =-\frac{2}{\pi}\oint_\gamma
 \St\left[
 \delta Y^\circ\mathsf T^\circ
 +Y^\circ\delta\mathsf T^\circ
 \right],
 \label{eq:variation-AdS-first-order-position}
\end{align}
where $\delta Y^\circ$ is given by
Eq.~\eqref{eq:right-position-variation}.  Combining this with the
matter terms in Eq.~\eqref{eq:right-bulk-boundary-variation-split}
gives
\begin{align}
 \left.\delta(S+I_{\rm pos})\right|_{\gamma,{\rm mat}}
 =\frac{2}{\pi}\oint_\gamma
 \St\left(
 \mathcal P^\bullet\Omega'^\bullet
 +\mathcal P^\circ\Omega'^\circ
 -\delta Y^\circ\mathsf T^\circ
 -Y^\circ\delta\mathsf T^\circ
 \right).
 \label{eq:bulk-plus-position-variation}
\end{align}
Variation of $\mathsf T^\circ$ first imposes
$Y^\circ|_\gamma=0$.  On this locus
Eq.~\eqref{eq:right-linear-position-variation} gives
$\delta Y^\circ=\Omega'^\circ$, and stationarity for arbitrary
$\Omega'^\bullet$, $\Omega'^\circ$, and $\delta\mathsf T^\circ$ yields
\begin{align}
 \mathcal P^\bullet\big|_\gamma&=0,
 &
 \mathsf T^\circ\big|_\gamma
 &=\mathcal P^\circ\big|_\gamma,
 &
 Y^\circ\big|_\gamma&=0.
 \label{eq:first-order-position-boundary-equations}
\end{align}
On the chosen logarithmic branch, the last equation implies
\begin{align}
 Y^\circ\big|_\gamma=0
 \quad\Longrightarrow\quad
 \mathcal C_R(g)\big|_\gamma=\mathbf 1_G
 \quad\Longrightarrow\quad
 R(g)\big|_\gamma=g\big|_\gamma.
 \label{eq:Y-zero-fixed-subgroup}
\end{align}
Here $\mathbf 1_G$ is the identity element of $G$.  Thus the boundary
representative belongs to $G^R$, and the reference brane is
$G^R/(H_0\cap G^R)$.
The second equation in
Eq.~\eqref{eq:first-order-position-boundary-equations} identifies
$\mathsf T^\circ$ as the anti-fixed canonical boundary momentum in the
Maurer--Cartan frame,
just as $\mathsf T_{A'}=\star dX_{A'}$ in flat space.

Independent ghost variations require a free ghost boundary term as
well.  For $r=\bullet,\circ$, define
\begin{align}
 \lambda_\pm^r=\bar\lambda^r\mathbin\pm\lambda^r,
 \qquad
 \omega_\pm^r=\bar\omega^r\mathbin\pm\omega^r.
 \label{eq:ghost-plus-minus}
\end{align}
The boundary polarization compatible with the bulk ghost symplectic
term is
\begin{align}
 I_{\rm gh}^{(0)}
 =-\frac{2}{\pi}\oint_\gamma d\sigma\,
 \St\left(
 \omega_-^\bullet\lambda_-^\bullet
 +\omega_+^\circ\lambda_+^\circ
 \right).
 \label{eq:free-ghost-boundary-action}
\end{align}
To state the constrained variational problem precisely, we use local
coordinates on each pure-spinor cone and choose local representatives
of the antighost gauge classes.  In this local description, independent
variation selects the gluing branch
\begin{align}
 \lambda_-^\bullet\big|_\gamma=0,
 \qquad
 \lambda_+^\circ\big|_\gamma=0,
 \qquad
 \omega_+^\bullet\big|_\gamma=0,
 \qquad
 \omega_-^\circ\big|_\gamma=0,
 \label{eq:free-ghost-gluing}
\end{align}
which is equivalent to Eq.~\eqref{ghostBCnew}.  The first two equations
are relations on the pure-spinor cones.  The last two are equalities of
antighost equivalence classes; the displayed zeroes refer to the chosen
local representatives.  Field-dependent ghost
interactions must be added to this free boundary action and determined
together with the matter interaction by ordinary variation and the
BRST-current condition of Section~\ref{sec:boundinteractions}.

For a translated brane the construction is applied to
$\widehat g=g_0^{-1}g$.  Outside the local logarithmic domain the
corresponding coset-level fixed-locus condition remains
\begin{align}\label{dirichletAdS}
  R(\widehat g)=\widehat g h,\qquad h\in H_0.
\end{align}
This is well defined because $R(H_0)=H_0$; for the reference
representative $g_0=\mathbf 1_G$, it reduces to $R(g)=gh$.

\subsection{Candidate half-BPS probe embeddings}
\label{sec:allowedDbranes}
We retain only Lorentzian boundary conditions with vanishing background
world-volume flux and a field-independent gluing involution.  A
Euclidean D$(-1)$-brane is therefore outside the domain of the paper,
not ruled out by it.  A half-BPS D9-brane is excluded within this domain
by the explicit fermionic gluing constraints summarized above; the fact
that no candidate sixteen-supercharge fixed algebra contains the full
bosonic isometry $\su(2,2)\oplus\su(4)$ is a symmetry cross-check rather
than a separate charge-state argument.

The subalgebras of $\psu(2,2|4)$ containing sixteen odd generators and
a maximal number of bosonic generators are classified
in~\cite{DHoker:2008wvd}.  Using the conventional labels of that
reference, the list is
\begin{itemize}
  \item $\mathfrak{su}(2|4)\oplus \su(2)$
  \item $\su(1,1|4)\oplus\su(1,1)$
  \item $\su(2,2|2)\oplus \su(2)$
  \item $\osp(4^*|4)$
  \item $\osp(4|4;\mathbb{R})$
  \item $\psu(2|2)\oplus\psu(2|2)\oplus \mathbb{R}^2$
  \item $\psu(1,1|2)\oplus\psu(1,1|2)\oplus \mathfrak{u}(1)^2$
  \item $\su(2|3)\oplus\su(2|1)$
  \item $\su(1,1|3)\oplus\su(1,1|1)$
  \item $\su(1,2|2)\oplus\su(1|2)$
\end{itemize}
We use these conventional algebra labels throughout.  For the two D3
cases, only the $\mathbb Z_4$ assignment of the Abelian directions will
be needed below.

Not every algebra in this list can be the fixed algebra of a
complex-linear, field-independent boundary involution compatible with
the sigma-model $\mathbb Z_4$ grading.  In particular, the unitary rows
that require a $3+1$ split of either defining four-dimensional space
fail the condition
\eqref{eq:R-z4-compatibility}, as we will explain in
Section~\ref{sec:matrix-involutions}. The compact
$\mathfrak{su}(2|4)\oplus\mathfrak{su}(2)$ row fails the same
compatibility condition for a different sign reason, also described
there.

This list was obtained as part of the classification of possible
symmetry superalgebras of half-BPS supergravity solutions.  It is a
necessary symmetry test here, not a classification of world-sheet
boundary involutions or probe dynamics: the converse implication is
explicitly absent in~\cite{DHoker:2008wvd}, and a single superalgebra
can act on inequivalent brane orbits.  In particular, every proposed
row must be tested against the independent chain
\begin{align}
 \begin{aligned}
  R&\ \rightarrow\ G^R\ \rightarrow\
  (\frakg_2^\bullet,\frakg_2^\circ)
  &\ \rightarrow\ \mathcal W=G^R/(G^R\cap H_0)
  \ \rightarrow\
  \text{boundary BRST invariance}.
 \end{aligned}
\end{align}
The representatives below pass the involution, ambient-real-form,
$\mathbb Z_4$-exchange, fixed-algebra, and grade-two-dimension tests.
The grade-exchange identity fixes the required exchange of the odd sectors, and
the automorphism property then preserves the pure-spinor cone.  The
zero-field boundary-BRST condition was proved directly above from the
fixed/anti-fixed decomposition.  We determine the isotropy
intersections and the physical D3 Abelian directions below.
Candidate fixed algebras associated with known probe configurations include

\begin{itemize}
  \item $\su(1,1|4)\oplus\su(1,1)$
  \item $\su(2,2|2)\oplus \su(2)$
  \item $\osp(4^*|4)$
  \item $\osp(4|4;\mathbb{R})$
  \item $\psu(1,1|2)\oplus\psu(1,1|2)\oplus \mathfrak{u}(1)$
  \item $\psu(2|2)\oplus\psu(2|2)\oplus \mathbb{R}$
\end{itemize}
We now summarize the corresponding zero-flux probe geometries
\cite{DHoker:2008wvd,Hanazawa:2016lvo}.  As we will describe in
Section~\ref{sec:matrix-involutions}, explicit inner and outer matrix
representatives realize these sectors and determine their grade-two
fixed spaces.  Together, these representatives realize all known
Lorentzian zero-flux D1, D3, D5, and D7 probe sectors considered in
the present framework.

\paragraph{$D1$-brane}
The zero-flux Lorentzian D1 embedding retained here has world-volume
$AdS_2$ and preserved supergroup $OSp(4^*|4)$. Its bosonic subgroup is
$SO(2,1)\times SO(3)\times SO(5)$: the first factor acts on $AdS_2$,
while $SO(3)\times SO(5)$ is the transverse rotational stabilizer. The
world-volume supercoset is
$OSp(4^*|4)/(SO(1,1)\times SO(3)\times SO(5))$.

\paragraph{$D3$-branes}
The two familiar D3 geometries are described by the conventional
labels $\psu(1,1|2)\oplus\psu(1,1|2)\oplus\mathfrak u(1)$ and
$\psu(2|2)\oplus\psu(2|2)\oplus\mathbb R$.  For the explicit diagonal
$2|2+2|2$ representatives, the $\mathbb Z_4$ grading places one
Abelian direction in $\frakg_0^\bullet$ and the other in
$\frakg_2^\bullet$.  In the $AdS_3\times S^1$ case, the former belongs
to the isotropy algebra and the latter generates the physical $S^1$.
For the giant-graviton-type configuration, the grade-zero direction
belongs to the sphere stabilizer and the grade-two direction generates
time.  The corresponding bosonic world-volumes are
$AdS_3\times S^1$ and $\mathbb R\times S^3$, respectively.  We verify
these grade assignments in Section~\ref{sec:matrix-involutions}.

\paragraph{$D5$-brane}
For the D5-brane there are two zero-flux possibilities. The first has
an $AdS_4\times S^2$ world-volume and supercoset
$OSp(4|4;\mathbb{R})/(SO(1,3)\times SO(2)\times SO(3))$. The second has
the same preserved algebra as the D1-brane, but its supercoset is
$OSp(4^*|4)/(SO(1,1)\times SO(3)\times SO(4))$ and its bosonic
world-volume is $AdS_2\times S^4$.

\paragraph{$D7$-brane}
The two probe D7 possibilities are
$SU(2,2|2)\times SU(2)$ and $SU(1,1|4)\times SU(1,1)$. The first
is a brane in $AdS_5\times S^3$ and the coset is given by
$(SU(2,2|2)\times SU(2))/(SO(1,4)\times SO(3)\times SO(2))$. The
other possibility is
$(SU(1,1|4)\times SU(1,1))/(SO(1,2)\times SO(5)\times SO(2))$, with
bosonic world-volume $AdS_3\times S^5$. These are probe statements;
the existence and asymptotics of fully backreacted D7 solutions require
separate analysis~\cite{DHoker:2008wvd}.

The isotropy groups can now be determined directly rather than inferred from the tangent dimensions alone. At the Lie-algebra level,
\begin{align}
 \operatorname{Lie}(G^R\cap H_0)
 =\frakg^\bullet\cap\frakg_0=\frakg_0^\bullet.
 \label{eq:fixed-isotropy-algebra}
\end{align}
For the identity components of the retained representatives we find
\begin{align}
\begin{array}{c|c}
\mathcal W_{\mathrm{bos}}&(G^R\cap H_0)_0\\ \hline
AdS_2&SO(1,1)\times SO(3)\times SO(5)\\
AdS_3\times S^1&SO(1,2)\times SO(4)\times U(1)_{\mathrm{stab}}\\
\mathbb R\times S^3&(SU(2))^3\times U(1)_{\mathrm{stab}}\\
AdS_2\times S^4&SO(1,1)\times SO(3)\times SO(4)\\
AdS_4\times S^2&SO(1,3)\times SO(2)\times SO(3)\\
AdS_5\times S^3&SO(1,4)\times SO(3)\times SO(2)\\
AdS_3\times S^5&SO(1,2)\times SO(5)\times SO(2)
\end{array}
\label{eq:probe-isotropy-table}
\end{align}
Here and below we suppress finite centers and global-cover choices.
In either D3 row, $U(1)_{\mathrm{stab}}$ denotes the one-parameter
subgroup generated by the grade-zero Abelian direction.  The
complementary grade-two direction supplies the $S^1$ or time direction
of the brane.

\section{\texorpdfstring{$D$-branes from $\mathbb Z_4$ symmetry}{D-branes from Z4 symmetry}}
\label{sec:matrix-involutions}

We now return to the compatibility between the boundary involution
$R$ and the automorphism $\Sigma$ generating the $\mathbb Z_4$ grading
of $\psu(2,2|4)$, reviewed in Appendix~\ref{appendix:psu}. For a
complex-linear $R$, the global condition from \eqref{ghostgluing1} is 
\begin{align}
  R\Sigma R^{-1}=\Sigma^{-1}.
  \label{eq:R-z4-compatibility}
\end{align}
This condition implies that $R$ preserves the even grades and exchanges $\mathfrak g_1$ with $\mathfrak g_3$. On the odd subspace, the compatibility condition reduces to $R\Sigma=-\Sigma R$. Once grade exchange is assumed, however, this
odd-sector relation is automatic and does not classify the admissible
maps.  We now give matrix representatives for all the unitary and
orthosymplectic rows retained above.  The common matrix identities are
collected in Appendix~\ref{appendix:matrix-involutions}.

\subsection{Diagonal inner representatives}

We use $[M]$ for the projective class of
$M\in\mathfrak{sl}(4|4;\mathbb C)$ modulo the central identity. Our
supertranspose convention is
\begin{align}
 M=\begin{pmatrix}A&\Psi\\ \Theta&B\end{pmatrix},
 \qquad
 M^\ST=\begin{pmatrix}
 A^\T&\Theta^\T\\ -\Psi^\T&B^\T
 \end{pmatrix}
 \label{eq:main-supertranspose-convention}.
\end{align}
Further details are collected in
Appendix~\ref{appendix:psu}.

We write the implementing matrix and the corresponding automorphism as
\begin{align}
 U&=\operatorname{diag}(U_{\mathrm{AdS}},U_S),&
 R_U([M])&=[UMU^{-1}].
 \label{eq:inner-sign-map}
\end{align}
This distinction is useful: $U$ is a matrix, whereas $R_U$ is the
involution of the algebra.  In the basis
$I_{2,2}=\operatorname{diag}(+,+,-,-)$, we define
\begin{align}
 u_0&={\bf 1}_4,&
 u_{\mathrm{a}}&=\operatorname{diag}(+,-,+,-),&
 u_{\mathrm{s}}&=\operatorname{diag}(+,+,-,-).
 \label{eq:three-sign-matrices}
\end{align}
The alternating matrix $u_{\mathrm{a}}$ splits the $SU(2,2)$ defining
space into two subspaces of signature $(1,1)$ and hence produces two
$\su(1,1)$ factors.  The aligned matrix $u_{\mathrm{s}}$ splits it into
definite subspaces and produces two compact $\su(2)$ factors.

For the $\mathbb Z_4$ test, let
\begin{align}
 j&=\begin{pmatrix}0&1\\-1&0\end{pmatrix},&
 J_0&=j\oplus j,&
 C_0&=\operatorname{diag}(J_0,J_0),
 \label{eq:z4-matrix-basis}\\
 \Sigma([M])&=[-C_0^{-1}M^\ST C_0].
\end{align}
Here $\Sigma$ acts on the complexified algebra; complex conjugation may exchange the $\mathrm i$ and $-\mathrm i$ eigenspaces of $\Sigma$. Real-form preservation is a separate condition on $R$. Conjugating the supertranspose map in Eq. \eqref{eq:z4-matrix-basis} shows that the compatibility relation $R_U\Sigma R_U^{-1}=\Sigma^{-1}$ holds precisely when
\begin{align}
 U_{\mathrm{AdS}}^{\T}J_0U_{\mathrm{AdS}}&=\tau J_0,&
 U_S^{\T}J_0U_S&=-\tau J_0,\qquad \tau=\pm1.
 \label{eq:inner-z4-sign-test}
\end{align}

This condition gives a basis-independent obstruction to the unitary
rows built from a $3+1$ split.  Let $E_+$ and $E_-$ be the eigenspaces
of an involution $U$ on a four-dimensional space equipped with the
non-degenerate skew form $J_0$.  If
$U^{\T}J_0U=J_0$, the two eigenspaces are symplectically orthogonal and
the restriction of $J_0$ to each one is non-degenerate.  Their
dimensions must therefore be even.  If instead
$U^{\T}J_0U=-J_0$, each eigenspace is isotropic; in four dimensions
both must then have dimension two.  Thus an involution with eigenspace
dimensions $(3,1)$ satisfies neither sign allowed in
Eq.~\eqref{eq:inner-z4-sign-test}.

For example, a fixed algebra
$\su(2|3)\oplus\su(2|1)$ requires a $2+2$ split of the $SU(2,2)$
defining space and a $3+1$ split of the $SU(4)$ defining space.  In the
canonical basis the latter may be represented by
\begin{align}
 u_{3,1}&=\operatorname{diag}(+,+,+,-),&
 u_{3,1}^{\T}J_0u_{3,1}&=j\oplus(-j)\not\propto J_0.
 \label{eq:three-one-obstruction}
\end{align}
No change of basis can remove this obstruction because it follows from
the eigenspace dimensions.  The outer family discussed below does not
provide an alternative: its fixed algebra is orthosymplectic rather
than a direct sum of unitary superalgebras.  
Hence, within the complex-linear, field-independent involutions considered here, no $R$ satisfying Eq. $\eqref{eq:R-z4-compatibility}$ has fixed algebra $\mathfrak{su}(2|3)\oplus\mathfrak{su}(2|1)$. The same argument excludes
$\su(1,1|3)\oplus\su(1,1|1)$ and
$\su(1,2|2)\oplus\su(1|2)$, with the $3+1$ split occurring in the
sphere and $AdS$ blocks, respectively.

The omitted compact row
$\mathfrak{su}(2|4)\oplus\mathfrak{su}(2)$ is excluded independently
of the chosen matrix basis.  Preserving the full $\mathfrak{su}(4)$ in
the sphere block forces $U_S$ to be scalar and hence
\begin{align}
 U_S^{\T}J_0U_S\propto +J_0.
\end{align}
On the $AdS$ block, the compact
$\mathfrak{su}(2)\oplus\mathfrak{su}(2)$ centralizer requires the two
eigenspaces of $U_{\mathrm{AdS}}$ to be definite with respect to the
Hermitian form of signature $(2,2)$.  A negative symplectic sign would
instead make both eigenspaces $J_0$-Lagrangian; in the
$\mathfrak{su}(2,2)$ real form such planes have split signature
$(1,1)$ and produce non-compact $\mathfrak{su}(1,1)$ factors.  The
compact split therefore also obeys
\begin{align}
 U_{\mathrm{AdS}}^{\T}J_0U_{\mathrm{AdS}}\propto +J_0.
\end{align}
The two blocks consequently have the same sign, whereas
Eq.~\eqref{eq:inner-z4-sign-test} requires opposite signs.  Thus no
real-form-preserving inner representative with this fixed algebra
satisfies Eq.~\eqref{eq:R-z4-compatibility}; the outer family has
orthosymplectic fixed algebras and provides no alternative.

The resulting representatives and their fixed grade-two dimensions are
\begin{align}
\begin{array}{c|c|c|c}
(U_{\mathrm{AdS}},U_S)&\text{fixed algebra}&
(\dim\frakg_{2,\mathrm{AdS}}^\bullet,\dim\frakg_{2,S}^\bullet)&
\text{probe body}\\ \hline
(u_{\mathrm{a}},u_0)&\su(1,1|4)\oplus\su(1,1)&(3,5)&AdS_3\times S^5\\
(u_0,u_{\mathrm{a}})&\su(2,2|2)\oplus\su(2)&(5,3)&AdS_5\times S^3\\
(u_{\mathrm{a}},u_{\mathrm{s}})
 &\psu(1,1|2)\oplus\psu(1,1|2)\oplus\mathfrak u(1)
 &(3,1)&AdS_3\times S^1\\
(u_{\mathrm{s}},u_{\mathrm{a}})
 &\psu(2|2)\oplus\psu(2|2)\oplus\mathbb R
 &(1,3)&\mathbb R\times S^3
\end{array}
\label{eq:unitary-involution-table}
\end{align}
All four rows have sixteen real fixed odd generators.  For the D3 rows,
the only additional information we need is the grade of each Abelian
block generator.  Define
\begin{align}
 z_{\mathrm{AdS}}
 &=
 \im\left[\operatorname{diag}(U_{\mathrm{AdS}},0)\right],
 &
 z_S
 &=
 \im\left[\operatorname{diag}(0,U_S)\right].
\end{align}
For $u=u_{\mathrm a}$ or $u=u_{\mathrm s}$,
\begin{align}
 u^{\T}J_0u=sJ_0,\qquad s=\pm1,
 \quad\Longrightarrow\quad
 \Sigma(u)=-J_0^{-1}u^{\T}J_0=-s u.
 \label{eq:D3-abelian-grades}
\end{align}
Since $u_{\mathrm a}^{\mathrm T}J_0u_{\mathrm a}=-J_0$, whereas $u_{\mathrm s}^{\mathrm T}J_0u_{\mathrm s}=J_0$, a block generator constructed from $u_{\mathrm a}$ has grade zero, whereas one constructed from $u_{\mathrm s}$ has grade two. For the \(AdS_3\times S^1\) row, $z_{\mathrm{AdS}}\in\mathfrak g_0^\bullet$, while $z_S\in\mathfrak g_2^\bullet$. For the $\mathbb R\times S^3$ row, $z_S\in\mathfrak g_0^\bullet$, while $z_{\mathrm{AdS}}\in\mathfrak g_2^\bullet$. In each case, the grade-zero direction belongs to the isotropy group in Eq. \eqref{eq:probe-isotropy-table}, whereas the grade-two direction is the physical $S^1$ or time direction.

\subsection{Outer orthosymplectic representatives}

No diagonal inner centralizer equals $\osp(4^*|4)$ or
$\osp(4|4;\mathbb R)$. The required orthosymplectic construction uses the complex-linear outer automorphism
\begin{align}
 R_C([M])=[-C^{-1}M^\ST C].
 \label{eq:outer-boundary-map}
\end{align}
Its fixed equation $M^\ST C+CM=0$ is orthosymplectic, and
\begin{align}
 R_C^2&=\operatorname{Ad}_{C^{-1}C^\ST\Pi},&
 \Pi&=\operatorname{diag}({\bf1}_4,-{\bf1}_4).
 \label{eq:outer-involution-square-main}
\end{align}
Thus $C^\ST=\epsilon C\Pi$, $\epsilon=\pm1$, is a sufficient
involution condition.

Let
\begin{align}
 J_1&=\begin{pmatrix}0&{\bf1}_2\\-{\bf1}_2&0\end{pmatrix},&
 H&=\begin{pmatrix}0&{\bf1}_2\\ {\bf1}_2&0\end{pmatrix}.
\end{align}
In the block order $\su(2,2)\oplus\su(4)$, convenient representatives
are
\begin{align}
 C_{\mathrm{D1}}&=\operatorname{diag}(-\im H,J_0),\nonumber\\
 C_{\mathrm{D5}(2,4)}&=\operatorname{diag}(-\im H,J_1),\nonumber\\
 C_{\mathrm{D5}(4,2)}&=\operatorname{diag}(\im J_1,{\bf1}_4).
 \label{eq:three-osp-C}
\end{align}
The associated automorphisms commute with the ambient semilinear
reality map
\begin{align}
 \phi_L(M)=-L_{\rm amb}^{-1}(M^*)^\ST L_{\rm amb},\qquad
 L_{\rm amb}=\operatorname{diag}(I_{2,2},-\im{\bf1}_4),
 \label{eq:ambient-real-map-main}
\end{align}
and satisfy the outer grade-exchange condition
\begin{align}
 C^\ST(C_0^\ST)^{-1}C\ \propto\ C_0^\ST\Pi.
 \label{eq:outer-z4-test-main}
\end{align}
Their real fixed algebras and tangent dimensions are
\begin{align}
\begin{array}{c|c|c|c}
C&\text{fixed real algebra}&
(\dim\frakg_{2,\mathrm{AdS}}^\bullet,\dim\frakg_{2,S}^\bullet)&
\text{probe body}\\ \hline
C_{\mathrm{D1}}&\osp(4^*|4)&(2,0)&AdS_2\\
C_{\mathrm{D5}(2,4)}&\osp(4^*|4)&(2,4)&AdS_2\times S^4\\
C_{\mathrm{D5}(4,2)}&\osp(4|4;\mathbb R)&(4,2)&AdS_4\times S^2
\end{array}
\label{eq:osp-involution-table}
\end{align}
The replacement $J_0\to J_1$ leaves the abstract
$\mathfrak{usp}(4)$ sphere algebra unchanged but changes its position
relative to the isotropy algebra.  This is why the same
$\osp(4^*|4)$ can describe the D1 and the $AdS_2\times S^4$ D5 with
different grade-two spaces and stabilizers.

Supertransposition and complex conjugation play distinct roles.  The
complex fixed algebra of $R_C$ is $\osp(4|4;\mathbb C)$, while the
physical real algebra is
\begin{align}
 \operatorname{Fix}(R_C)\cap\operatorname{Fix}(\phi_L).
\end{align}
On $\psu(2,2|4)$ the same map can be rewritten in an antilinear-looking
form $R_C(M)=D M^*D^{-1}$ with
$D=C^{-1}\overline{L}_{\rm amb}$. A standalone
map $-C^{-1}(M^*)^\ST C$, however, fixes a real form of the entire
$\mathfrak{psl}(4|4;\mathbb C)$ rather than an orthosymplectic
subalgebra.  Moreover, an antilinear map already conjugates
$\im^k\to\im^{-k}$, so its grade-reversal equation is not the same as
the complex-linear equation used above.

The matrix classification stops at the field-independent,
zero-background gluing problem stated in the Introduction.  No
flux-deformed branches are considered in this paper.

\section{Boundary interactions in \texorpdfstring{$AdS$}{AdS}}
\label{sec:boundinteractions}

Let $\mathcal W=G^R/(G^R\cap H_0)$ denote the reference brane orbit; a
translated representative is obtained by the conjugation prescription
given above.  The grade-two spaces $\frakg_2^\bullet$ and
$\frakg_2^\circ$ determine, respectively, the tangent and normal
directions of the bosonic body of $\mathcal W$.  This statement concerns only the bosonic geometry and does not restrict the fermionic grades that can enter the boundary interaction.

\subsection{Matter and ghost interactions}

The field $g_\bullet$ is the local representative of the reference
brane introduced in Eq.~\eqref{eq:local-fixed-normal-factorization}.
Let $\mathcal A$ be an Abelian one-form superconnection on the
preserved boundary superspace $\mathcal W$.  We denote its
Maurer--Cartan-frame representative by
$\mathbf A^\bullet(g_\bullet)$, defined through
\begin{align}
 \mathcal A(V)
 =\St\left(V^\bullet\mathbf A^\bullet\right),
 \qquad V\in T\mathcal W,
 \label{eq:open-superconnection-frame}
\end{align}
where
$\mathfrak m^{\bullet,\circ}:=\mathfrak m\cap
\frakg^{\bullet,\circ}$ and
$V^\bullet\in\mathfrak m^\bullet$ is the
Maurer--Cartan-frame representative of $V$.
Let $\Phi^\circ(g_\bullet)$ be a section of the local anti-fixed coset
bundle. The fluctuating brane is the graph
\begin{align}
 g=g_\bullet e^{\Phi^\circ(g_\bullet)}.
 \label{eq:fluctuating-brane-graph}
\end{align}
Both frame fields are equivariant under the residual group
$H_0\cap G^R$: $\Phi^\circ$ and $\mathbf A^\bullet$ transform by the
adjoint action, so that $\mathcal A$ is an invariant one-form. We
reserve $A$ without boldface for the grade-zero isotropy connection in
Eq.~\eqref{eq:right-frame-current-identities}. Since
$\mathfrak m=\frakg_1\oplus\frakg_2\oplus\frakg_3$, the open
superfields contain bosonic grade-two components and fermionic
grade-one and grade-three components; they are not bosonic
truncations.
The Abelian open-string gauge transformation is
\begin{align}
 \mathcal A\longmapsto
 \mathcal A+d_{\mathcal W}\Lambda,
\end{align}
where $d_{\mathcal W}$ is the exterior derivative on the boundary
superspace.  This leaves the Wilson coupling invariant, and the frame
representative $\mathbf A^\bullet$ transforms accordingly.

Along a boundary path in the reference orbit, write
\begin{align}
 g_\bullet^{-1}\partial_\sigma g_\bullet
 =a_\sigma+V_\sigma^\bullet,
 \qquad
 a_\sigma\in\frakg_0^\bullet,\quad
 V_\sigma^\bullet\in\mathfrak m^\bullet.
 \label{eq:reference-boundary-velocity}
\end{align}
By Eq.~\eqref{eq:open-superconnection-frame}, the pullback of the
superconnection is then
$\mathcal A_\sigma=\St(V_\sigma^\bullet\mathbf A^\bullet)$.
The matter interaction deforms the reference position term
Eq.~\eqref{eq:AdS-first-order-position} by the embedding field and the
tangential superconnection:
\begin{align}
 I_{\rm mat}^{(1)}
 &=I_{\rm pos}
 +\frac{2}{\pi}\oint_\gamma d\sigma\,
 \St\left(V_\sigma^\bullet\mathbf A^\bullet\right)
 +\frac{2}{\pi}\oint_\gamma
 \St\left(\Phi^\circ\mathsf T^\circ\right)
 \nonumber\\
 &=\frac{2}{\pi}\left\{
 \oint_\gamma d\sigma\,
 \St\left(V_\sigma^\bullet\mathbf A^\bullet\right)
 -\oint_\gamma
 \St\left[(Y^\circ-\Phi^\circ)\mathsf T^\circ\right]
 \right\}.
 \label{eq:first-order-matter-interaction}
\end{align}
The Wilson coupling has been normalized with the same \(2/\pi\) as
the bulk boundary one-form and the position term; the ghost boundary
terms use the same convention.  This common factor consequently drops
out of the boundary equations.  Variation of
$\mathsf T^\circ$ replaces the reference condition $Y^\circ=0$ by
$Y^\circ=\Phi^\circ$.  The gauge coupling continues to use the
Maurer--Cartan velocity of the reference orbit.  Its relation to the
fixed component of the current on the displaced graph will be derived
below.

The four zero-field ghost conditions in
Eq.~\eqref{eq:free-ghost-gluing} must deform in the presence
of the open-string fields.  In the boundary polarization defined in
Eq.~\eqref{eq:ghost-plus-minus}, their quadratic field-dependent
deformation uses four even, ghost-number-zero linear couplings
\begin{align}
 \mathcal M_{rs}:
 \mathfrak m_{\bar 1}^{\,s}
 \longrightarrow
 \mathfrak m_{\bar 1}^{\,r},
 \qquad
 r,s\in\{\bullet,\circ\},
\end{align}
where
$\mathfrak m_{\bar 1}=\frakg_1\oplus\frakg_3$ and
$\mathfrak m_{\bar1}^{\,s}
=\mathfrak m_{\bar1}\cap\frakg^s$.
The first index
gives the output $R$-parity and the second the input $R$-parity.  We do
not assume at this stage how these couplings depend on
$\mathbf A^\bullet$ and $\Phi^\circ$; the interacting field equations
will constrain their values on the allowed boundary pure-spinor cone.
Residual-isotropy invariance requires
\begin{align}
 \left.\mathcal M_{rs}\right|_{g_\bullet h}
 \!\left(\operatorname{Ad}_{h^{-1}}v\right)
 =
 \operatorname{Ad}_{h^{-1}}
 \left[
 \left.\mathcal M_{rs}\right|_{g_\bullet}(v)
 \right],
 \qquad h\in H_0\cap G^R.
 \label{eq:M-residual-equivariance}
\end{align}
The required ghost interaction is
\begin{align}
 I_{\rm gh}^{\rm int}
 =-\frac{2}{\pi}\oint_\gamma d\sigma\,
 \St\Big[
 &\omega_-^\bullet
  \mathcal M_{\bullet\bullet}(\lambda_+^\bullet)
 +\omega_-^\bullet
  \mathcal M_{\bullet\circ}(\lambda_-^\circ)
 \nonumber\\[-2pt]
 &+\omega_+^\circ
  \mathcal M_{\circ\bullet}(\lambda_+^\bullet)
 +\omega_+^\circ
  \mathcal M_{\circ\circ}(\lambda_-^\circ)
 \Big].
 \label{eq:field-dependent-ghost-interaction}
\end{align}
The terms proportional to $\omega_-^\bullet$ and
$\omega_+^\circ$ deform the first two conditions in
Eq.~\eqref{eq:free-ghost-gluing}.  Variation of
$\lambda_+^\bullet$ and $\lambda_-^\circ$ gives the corresponding
deformation of the two antighost conditions.
The adjoint of each coupling is defined by
\begin{align}
 \St\left[
 w^r\mathcal M_{rs}(v^s)\right]
 =
 \St\left[
 \mathcal M_{rs}^\dagger(w^r)v^s\right],
 \qquad r,s\in\{\bullet,\circ\}
 \quad\text{(no sum)},
 \label{eq:componentwise-ghost-pairing}
\end{align}
for $v^s\in\mathfrak m_{\bar 1}^{\,s}$ and
$w^r\in\mathfrak m_{\bar 1}^{\,r}$.  We choose the branch continuously
connected to the free gluing.  Working in the same local coordinates
on the pure-spinor cones and the same local antighost gauge slice used
above, variation of
$I_{\rm gh}^{(0)}+I_{\rm gh}^{\rm int}$ gives the field-dependent
version of Eq.~\eqref{eq:free-ghost-gluing}:
\begin{align}
 \lambda_-^\bullet\big|_\gamma
 &=-\mathcal M_{\bullet\bullet}(\lambda_+^\bullet)
   -\mathcal M_{\bullet\circ}(\lambda_-^\circ),
 \nonumber\\
 \lambda_+^\circ\big|_\gamma
 &=-\mathcal M_{\circ\bullet}(\lambda_+^\bullet)
   -\mathcal M_{\circ\circ}(\lambda_-^\circ),
 \nonumber\\
 \omega_+^\bullet\big|_\gamma
 &=\mathcal M_{\bullet\bullet}^\dagger(\omega_-^\bullet)
   +\mathcal M_{\circ\bullet}^\dagger(\omega_+^\circ),
 \nonumber\\
 \omega_-^\circ\big|_\gamma
 &=\mathcal M_{\bullet\circ}^\dagger(\omega_-^\bullet)
   +\mathcal M_{\circ\circ}^\dagger(\omega_+^\circ).
 \label{eq:interacting-ghost-gluing}
\end{align}
Equation~\eqref{eq:interacting-ghost-gluing} is precisely the
field-dependent deformation of Eq.~\eqref{eq:free-ghost-gluing}.  Its
first two equations are relations on the allowed boundary pure-spinor
data.  The last two are equalities in the antighost
quotient, written here for the chosen local representatives.  To keep
the subsequent formulas short, we denote the two output combinations
by
\begin{align*}
 \mathcal M^\bullet
 \left(\lambda_+^\bullet,\lambda_-^\circ\right)
 &:=
 \mathcal M_{\bullet\bullet}(\lambda_+^\bullet)
 +\mathcal M_{\bullet\circ}(\lambda_-^\circ),
 \\
 \mathcal M^\circ
 \left(\lambda_+^\bullet,\lambda_-^\circ\right)
 &:=
 \mathcal M_{\circ\bullet}(\lambda_+^\bullet)
 +\mathcal M_{\circ\circ}(\lambda_-^\circ).
\end{align*}
These are shorthands, not additional couplings.  Inverting the
definitions in Eq.~\eqref{eq:ghost-plus-minus}, the first two equations
in Eq.~\eqref{eq:interacting-ghost-gluing} reconstruct the boundary
ghosts as
\begin{align}
 \left.\lambda\right|_\gamma
 &=\frac12\Big[
 \lambda_+^\bullet-\lambda_-^\circ
 +\mathcal M^\bullet
   \left(\lambda_+^\bullet,\lambda_-^\circ\right)
 -\mathcal M^\circ
   \left(\lambda_+^\bullet,\lambda_-^\circ\right)
 \Big],
 \nonumber\\
 \left.\bar\lambda\right|_\gamma
 &=\frac12\Big[
 \lambda_+^\bullet+\lambda_-^\circ
 -\mathcal M^\bullet
   \left(\lambda_+^\bullet,\lambda_-^\circ\right)
 -\mathcal M^\circ
   \left(\lambda_+^\bullet,\lambda_-^\circ\right)
 \Big].
 \label{eq:boundary-ghost-reconstruction}
\end{align}
At zero fields the allowed boundary data are inherited from the free
condition $\bar\lambda=R(\lambda)$:
\begin{align}
 \lambda_+^\bullet
 =\lambda+R(\lambda),
 \qquad
 \lambda_-^\circ
 =R(\lambda)-\lambda,
 \qquad
 \lambda\in\frakg_1,
 \qquad
 \{\lambda,\lambda\}=0.
\end{align}
The present ansatz keeps this domain fixed; a more general
field-dependent gluing could also deform it.  At nonzero fields the two
elements reconstructed in Eq.~\eqref{eq:boundary-ghost-reconstruction}
must remain in $\frakg_1$ and $\frakg_3$, respectively, and obey
$\{\lambda,\lambda\}=0$ and
$\{\bar\lambda,\bar\lambda\}=0$.  Thus the two output combinations of
the four $\mathcal M_{rs}$ are the fixed and anti-fixed components of
one grade-three deformation on the allowed boundary data.  This
is the meaning of grade and pure-spinor-cone preservation below; it is
not an additional boundary condition.

The adjoints must also be compatible with the antighost gauge quotient.
In superalgebra notation the transformations are
\begin{align}
 \delta_X\omega&=[X,\lambda],
 &X&\in\frakg_2,
 \nonumber\\
 \delta_{\bar X}\bar\omega&=[\bar X,\bar\lambda],
 &\bar X&\in\frakg_2.
 \label{eq:antighost-gauge-transformations}
\end{align}
Here $X$ and $\bar X$ are independent.  Locally we choose a
representative of the quotient by these shifts; all antighost
equalities and pairings below are understood on that slice.
As emphasized for pure-spinor boundary states in
Ref.~\cite{Schiappa:2005mk}, the ghost couplings and their boundary
conditions cannot be separated from this gauge symmetry: in the
flat-space analysis the complete action is invariant once the boundary
conditions are imposed.  We therefore require the ghost deformation
defined by the four $\mathcal M_{rs}$ to preserve this gauge
compatibility, as well as the pure-spinor cone.  We retain these
requirements as restrictions on the couplings, but do not solve them
in the present general formalism.

\subsection{Tangent vectors on the fluctuating brane}

The equation
$g=g_\bullet e^{\Phi^\circ(g_\bullet)}$ specifies the position of the
fluctuating brane.  To vary its boundary interaction, however, we must
also distinguish a motion on the reference orbit from the fixed
component of its lift to the displaced graph.  This distinction is
essential at finite $\Phi^\circ$, because the lift develops both
coset and grade-zero components.

Let $V\in T\mathcal W$ and write its body-fixed representative on the
reference orbit as
\begin{align}
 g_\bullet^{-1}\delta_Vg_\bullet
 =a_V+V^\bullet,
 \qquad
 a_V\in\frakg_0^\bullet,\quad
 V^\bullet\in\mathfrak m^\bullet.
 \label{eq:fixed-tangent-lift}
\end{align}
The residual-isotropy covariant derivative of the embedding is
\begin{align}
 \nabla_V\Phi^\circ
 :=\delta_V\Phi^\circ+[a_V,\Phi^\circ].
 \label{eq:embedding-covariant-directional-derivative}
\end{align}
Applying the product rule to
$g=g_\bullet e^{\Phi^\circ}$ gives the exact full Maurer--Cartan
variation
\begin{align}
 g^{-1}\delta_Vg
 &=a_V+\Gamma_\Phi(V^\bullet),
 \nonumber\\
 \Gamma_\Phi(V^\bullet)
 &:=
 e^{-\operatorname{ad}_{\Phi^\circ}}V^\bullet
 +\operatorname{dexp}_{-\Phi^\circ}
   \bigl(\nabla_V\Phi^\circ\bigr).
 \label{eq:full-graph-differential}
\end{align}
The connection $a_V$ remains fixed under $R$, but
$\Gamma_\Phi$ generally has induced grade-zero components of both
$R$-parities.  These components are part of the exact lift and must be
retained until the physical coset projection is taken.

To display the two parity blocks, set
$A_\Phi=\operatorname{ad}_{\Phi^\circ}$.  Before the
$\mathfrak m$ projection, Eq.~\eqref{eq:full-graph-differential} gives
\begin{align}
 P_\bullet\Gamma_\Phi(V^\bullet)
 &=
 \cosh(A_\Phi)V^\bullet
 +\frac{1-\cosh(A_\Phi)}{A_\Phi}\,
   \nabla_V\Phi^\circ,
 \nonumber\\
 P_\circ\Gamma_\Phi(V^\bullet)
 &=
 \frac{\sinh(A_\Phi)}{A_\Phi}
 \left(\nabla_V\Phi^\circ-A_\Phi V^\bullet\right).
 \label{eq:full-graph-parity-blocks}
\end{align}
The ratios denote their regular power series.  We now define the two
physical blocks of the graph differential by
\begin{align}
 \mathcal T_\Phi(V^\bullet)
 &:=
 P_{\mathfrak m}P_\bullet\Gamma_\Phi(V^\bullet),
 &
 \mathcal N_\Phi(V^\bullet)
 &:=
 P_{\mathfrak m}P_\circ\Gamma_\Phi(V^\bullet).
 \label{eq:physical-graph-blocks}
\end{align}
Thus $\mathcal T_\Phi$ is the fixed coset component of the lifted
tangent, while $\mathcal N_\Phi$ is its anti-fixed coset component.
At the reference brane,
\begin{align*}
 \mathcal T_0(V^\bullet)=V^\bullet,
 \qquad
 \mathcal N_0(V^\bullet)=\nabla_V\Phi^\circ.
\end{align*}
The first corrections are
\begin{align*}
 \mathcal T_\Phi(V^\bullet)
 &=P_{\mathfrak m}\left(
 V^\bullet-\frac12[\Phi^\circ,\nabla_V\Phi^\circ]
 +\frac12[\Phi^\circ,[\Phi^\circ,V^\bullet]]
 +\cdots\right),\\
 \mathcal N_\Phi(V^\bullet)
 &=P_{\mathfrak m}\left(
 \nabla_V\Phi^\circ-[\Phi^\circ,V^\bullet]
 +\frac16[\Phi^\circ,[\Phi^\circ,\nabla_V\Phi^\circ]]
 -\frac16[\Phi^\circ,[\Phi^\circ,[\Phi^\circ,V^\bullet]]]
 +\cdots\right).
\end{align*}
In particular, the anti-fixed grade-zero compensator beginning with
$-[\Phi^\circ,V^\bullet]$ is present in the full second line of
Eq.~\eqref{eq:full-graph-parity-blocks}; it is removed only by
$P_{\mathfrak m}$ after the full lift has been formed.

Since $\mathcal T_0$ is the identity, $\mathcal T_\Phi$ is locally
invertible near the reference brane.  It is useful to express the
normal response in the fixed frame of the displaced graph:
\begin{align}
 \Gamma_\Phi(V^\bullet)
 &=
 \mathcal T_\Phi(V^\bullet)
 +\mathcal N_\Phi(V^\bullet).
 \label{eq:graph-tangent-fixed-normal-split}
\end{align}
The actual fixed coset component of this tangent is
\begin{align}
 U^\bullet:=\mathcal T_\Phi(V^\bullet).
\end{align}
Therefore the reference-orbit vector associated with
$U^\bullet$ is
\begin{align}
 V^\bullet=\mathcal T_\Phi^{-1}(U^\bullet),
\end{align}
and the anti-fixed component of the same graph tangent is
\begin{align}
 \mathcal N_\Phi(V^\bullet)
 =
 \mathcal N_\Phi\left(
 \mathcal T_\Phi^{-1}(U^\bullet)\right).
\end{align}
This gives the nonlinear embedding operator
\begin{align}
 \mathcal D_\Phi
 :=\mathcal N_\Phi\circ\mathcal T_\Phi^{-1}.
 \label{eq:nonlinear-embedding-operator}
\end{align}
For an embedded fixed vector $U^\bullet$,
$\mathcal D_\Phi(U^\bullet)$ is therefore the anti-fixed coset
component of the same graph tangent.  This definition is exact,
contains the required isotropy compensator, and maps
$\mathfrak m^\bullet$ to $\mathfrak m^\circ$ by construction.

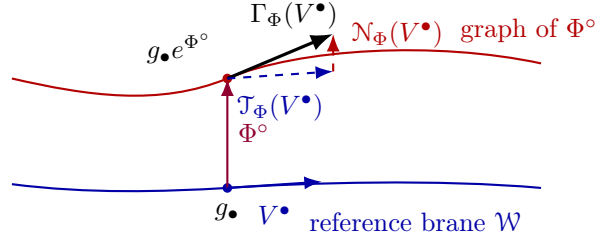
\begin{figure}[t]
\centering
\begin{tikzpicture}[
  >=Latex,
  font=\small,
  every node/.style={align=center}
]
  \coordinate (base) at (3.10,0.70);
  \coordinate (graph) at (3.10,2.15);
  \draw[thick,blue!65!black]
    (0.25,0.75)
    .. controls (1.25,0.62) and (2.25,0.65) .. (base)
    .. controls (4.15,0.77) and (5.90,0.82) .. (7.25,0.70);
  \draw[thick,red!70!black]
    (0.25,2.15)
    .. controls (1.25,1.92) and (2.20,1.78) .. (graph)
    .. controls (4.05,2.55) and (5.85,2.65) .. (7.25,2.35);
  \node[blue!65!black,anchor=east] at (7.12,0.25)
    {reference brane $\mathcal W$};
  \node[red!70!black,anchor=west] at (6.05,2.78)
    {graph of $\Phi^\circ$};

  \fill[blue!65!black] (base) circle (1.8pt);
  \fill[red!70!black] (graph) circle (1.8pt);
  \node[below=2pt of base] {$g_\bullet$};
  \node[above left=3pt of graph] {$g_\bullet e^{\Phi^\circ}$};
  \draw[->,purple!75!black,thick] (base) -- (graph)
    node[midway,right] {$\Phi^\circ$};

  \draw[->,blue!65!black,thick] (base) -- ++(1.25,0.083)
    node[midway,below=4pt] {$V^\bullet$};
  \draw[->,black,very thick] (graph) -- ++(1.40,0.59)
    node[pos=0.63,above=4pt] {$\Gamma_\Phi(V^\bullet)$};
  \draw[->,blue!65!black,dashed,thick] (graph) -- ++(1.40,0.093)
    node[midway,below=3pt] {$\mathcal T_\Phi(V^\bullet)$};
  \draw[->,red!70!black,dashed,thick]
    ($(graph)+(1.40,0.093)$) -- ($(graph)+(1.40,0.59)$)
    node[at end,right=3pt] {$\mathcal N_\Phi(V^\bullet)$};
\end{tikzpicture}
\caption{Local geometry of the nonlinear embedding operator.  The
drawing is schematic: fixed and anti-fixed denote $R$-parity,
represented here by their tangent and transverse interpretation near
the reference brane.  The full lift contains additional grade-zero
components not shown in the drawing.}
\label{fig:embedding-lift-operator}
\end{figure}

The comparison with Berkovits--Pershin is most transparent in
components.  Along the preserved fermionic directions,
$\Phi^\circ$ plays the role of their spinor superfield $W$, so that
$D\Phi^\circ$ corresponds schematically to $DW$.  The linearized graph
derivative is the covariant combination
$\nabla_V\Phi^\circ-[\Phi^\circ,V^\bullet]$, and
Eq.~\eqref{eq:nonlinear-embedding-operator} adds terms cubic and higher
in the fluctuating fields.  The integrated-vertex analysis in
$AdS_5\times S^5$ provides a useful analogue for the mixing involved:
its superfield chain contains the combination
$\nabla W-\tfrac12\eta\nabla A$, rather than $\nabla W$ alone,
where $\eta^{\alpha\bar\beta}$ is the covariantly constant RR
bispinor used in the conventions of Appendix~\ref{appendix:psu}
\cite{Chandia:2017afc}.  This motivates organizing the boundary
variables through the fermionic components of
$\mathbf A^\bullet$ and $\Phi^\circ$, without assuming that the
closed-string vertex argument alone determines the present boundary
ansatz.
Suppressing the index contractions and superalgebra brackets, the odd
part of the operator therefore has the schematic expansion
\begin{align}
 \mathcal D_\Phi
 \sim D\Phi+\Phi^2D\Phi+\cdots.
 \label{eq:embedding-operator-schematic}
\end{align}
With $W\sim\Phi$, this is the same schematic
$DW+W^2DW+\cdots$ structure found in the flat-space boundary
conditions of Ref.~\cite{Berkovits:2002ag}. The nested brackets retain
the precise curved-supercoset ordering; their higher powers resum the
additional nonlinearities generated by the nonlinear supercoset
embedding geometry.
We define its supertrace adjoint by
\begin{align*}
 \St\left[
 W^\circ\mathcal D_\Phi(V^\bullet)\right]
 =
 \St\left[
 V^\bullet\mathcal D_\Phi^\dagger(W^\circ)\right],
 \qquad
 V^\bullet\in\mathfrak m^\bullet,\quad
 W^\circ\in\mathfrak m^\circ.
\end{align*}

At the Abelian level the geometric open curvature is
$\mathcal F=d_{\mathcal W}\mathcal A$. We denote its
Maurer--Cartan-frame representative on the reference orbit by
$\mathbf F$.  For homogeneous tangent vectors its meaning is
\begin{align}
 \mathbf F(V,W)
 &=
 V\!\left[\mathcal A(W)\right]
 -(-1)^{|V||W|}
 W\!\left[\mathcal A(V)\right]
 -\mathcal A([V,W]).
 \label{eq:Abelian-open-curvature}
\end{align}
Thus the isotropy connection and the frame-anholonomy, or equivalently
the supertorsion, are already included.  At the Abelian level there is
no $\mathbf A^\bullet\wedge\mathbf A^\bullet$ term.  The curvature force
along the reference boundary path is the covector
\begin{align}
 \St\left(
 V^\bullet\mathbf F_\sigma^\bullet\right)
 &:=
 \mathbf F\left(V^\bullet,V_\sigma^\bullet\right),
 &
 \mathbf F_\sigma^\bullet
 &=-\iota_{V_\sigma^\bullet}\mathbf F.
 \label{eq:open-curvature-force-covector}
\end{align}
The second equality follows from graded antisymmetry because the
boundary tangent $V_\sigma^\bullet$ is even.  With this convention the
Wilson-line variation is
\begin{align}
 \delta\oint_\gamma d\sigma\,
 \St\left(V_\sigma^\bullet\mathbf A^\bullet\right)
 =\oint_\gamma d\sigma\,
 \St\left(V^\bullet\mathbf F_\sigma^\bullet\right).
 \label{eq:open-curvature-force}
\end{align}
For use in the BRST calculation, we pull tensors to the fixed frame of
the displaced graph.  For $U,W\in\mathfrak m^\bullet$, we define
\begin{align}
 \mathbf F_\Phi(U,W)
 &:=
 \mathbf F\left(
 \mathcal T_\Phi^{-1}U,
 \mathcal T_\Phi^{-1}W\right),
 &
 \nabla_U^\Phi
 &:=
 \nabla_{\mathcal T_\Phi^{-1}U}.
 \label{eq:embedded-frame-pullbacks}
\end{align}
If $U_\sigma^\bullet=\mathcal T_\Phi(V_\sigma^\bullet)$, then
\begin{align}
 \St\left(U^\bullet\mathbf F_{\Phi,\sigma}^\bullet\right)
 &:=
 \mathbf F_\Phi(U^\bullet,U_\sigma^\bullet),
 &
 \mathbf F_{\Phi,\sigma}^\bullet
 &:=-\iota_{U_\sigma^\bullet}\mathbf F_\Phi.
 \label{eq:embedded-curvature-force}
\end{align}
Our contraction convention is
$(\iota_U\mathbf F_\Phi)(W)=\mathbf F_\Phi(U,W)$; the minus sign in
\eqref{eq:embedded-curvature-force} therefore places the even boundary
velocity in the second slot of the force covector.

After imposing the ghost equations in
Eq.~\eqref{eq:interacting-ghost-gluing}, we introduce the independent
covariant normal variation
\begin{align*}
 \Xi^\circ:=\delta Y^\circ+[a_V,Y^\circ]
 \in\mathfrak m^\circ.
\end{align*}
For the auxiliary one-form we likewise use the covariant variation
\begin{align*}
 \Delta\mathsf T^\circ
 :=\delta\mathsf T^\circ+[a_V,\mathsf T^\circ].
\end{align*}
The two connection terms cancel inside the supertrace, so rewriting the
position coupling with $\Xi^\circ$, $\nabla_V\Phi^\circ$, and
$\Delta\mathsf T^\circ$ is exactly the ordinary off-shell variation.
The exact product rule gives
\begin{align*}
 g^{-1}\delta g
 =a_V+e^{-\operatorname{ad}_{Y^\circ}}V^\bullet
 +\operatorname{dexp}_{-Y^\circ}(\Xi^\circ).
\end{align*}
Here $\nabla_{V^\bullet}\mathcal M^r$ denotes the
residual-isotropy covariant variation of the corresponding output
combination along the reference-orbit vector $V^\bullet$, with its
pure-spinor arguments held fixed.  It is the induced covariant
derivative on the relevant
$\operatorname{Hom}(\mathfrak m_{\bar1}^{\,s},
\mathfrak m_{\bar1}^{\,r})$ bundle and therefore acts on both its input
and output frame indices.
The remaining ordinary boundary variation is therefore
\begin{align}
 \left.\delta\left(
 S+I_{\rm mat}^{(1)}+I_{\rm gh}^{(0)}
 +I_{\rm gh}^{\rm int}\right)\right|_{\gamma}
 =\frac{2}{\pi}\oint_\gamma d\sigma\,\Bigg\{&
 \St\Big[
 \mathcal P\left(
 e^{-\operatorname{ad}_{Y^\circ}}V^\bullet
 +\operatorname{dexp}_{-Y^\circ}(\Xi^\circ)\right)
 \nonumber\\
 &+V^\bullet\mathbf F_\sigma^\bullet
 -\left(\Xi^\circ-\nabla_V\Phi^\circ\right)
  \mathsf T^\circ
 \nonumber\\
 &-(Y^\circ-\Phi^\circ)\Delta\mathsf T^\circ
 \Big]
 \nonumber\\
 &-\St\Big[
 \omega_-^\bullet
 \left(\nabla_{V^\bullet}\mathcal M^\bullet\right)
 \left(\lambda_+^\bullet,\lambda_-^\circ\right)
 \nonumber\\
 &\hspace{18mm}
 +\omega_+^\circ
 \left(\nabla_{V^\bullet}\mathcal M^\circ\right)
 \left(\lambda_+^\bullet,\lambda_-^\circ\right)
 \Big]
 \Bigg\}.
 \label{eq:complete-interacting-boundary-variation}
\end{align}
At this stage $\Delta\mathsf T^\circ$, $\Xi^\circ$, and $V^\bullet$
are independent.  The open superfields are background fields on
$\mathcal W$; their changes in this world-sheet variation are induced
only by $V^\bullet$.  The coefficient of
$\Delta\mathsf T^\circ$ gives the graph condition
\begin{align}
 Y^\circ=\Phi^\circ.
 \label{eq:interacting-graph-condition}
\end{align}
We impose this equation only after taking the variation.

Once Eq.~\eqref{eq:interacting-graph-condition} has been imposed, the
coefficient of the independent normal coordinate variation gives the
normal momentum equation
\begin{align}
 \mathsf T^\circ
 =
 P_{\mathfrak m}P_\circ
 \left(\operatorname{dexp}_{-\Phi^\circ}\right)^\dagger
 \mathcal P.
 \label{eq:interacting-normal-momentum}
\end{align}
Since
$(\operatorname{ad}_{\Phi^\circ})^\dagger
=-\operatorname{ad}_{\Phi^\circ}$, we have
$(\operatorname{dexp}_{-\Phi^\circ})^\dagger
=\operatorname{dexp}_{\Phi^\circ}$.  Thus the equation determines the
auxiliary field and reduces to
$\mathsf T^\circ=\mathcal P^\circ$ at $\Phi^\circ=0$.
It also gives the exact identity
\begin{align*}
 \St\left(\nabla_V\Phi^\circ\mathsf T^\circ\right)
 =
 \St\left[
 \mathcal P\,
 \operatorname{dexp}_{-\Phi^\circ}
 \left(\nabla_V\Phi^\circ\right)\right].
\end{align*}
Together with the first term in
Eq.~\eqref{eq:complete-interacting-boundary-variation}, this reconstructs
the full graph differential $\Gamma_\Phi$.  Introduce the embedded
fixed vector
$U^\bullet=\mathcal T_\Phi(V^\bullet)$.  The complete tangential
boundary condition is then
\begin{align}
 0={}&\St\left[
 U^\bullet\left(
 \mathcal P^\bullet+\mathbf F_{\Phi,\sigma}^\bullet\right)
 +\mathcal D_\Phi(U^\bullet)\mathcal P^\circ
 \right]
 \nonumber\\
 &-\St\Bigg[
 \omega_-^\bullet
 \left(\nabla_{U^\bullet}^\Phi\mathcal M^\bullet\right)
 \left(\lambda_+^\bullet,\lambda_-^\circ\right)
 \nonumber\\
 &\hspace{14mm}
 +\omega_+^\circ
 \left(\nabla_{U^\bullet}^\Phi\mathcal M^\circ\right)
 \left(\lambda_+^\bullet,\lambda_-^\circ\right)
 \Bigg],
 \qquad\text{for every }U^\bullet.
 \label{eq:interacting-fixed-momentum}
\end{align}
We keep this equation in paired form because $U^\bullet$ is arbitrary; no pure-spinor direction has been selected in deriving it. Equations \eqref{eq:interacting-ghost-gluing}, \eqref{eq:interacting-graph-condition}, \eqref{eq:interacting-normal-momentum}, and \eqref{eq:interacting-fixed-momentum} exhaust the independent variations in Eq. \eqref{eq:complete-interacting-boundary-variation}. Consequently, the boundary one-form in that equation vanishes. In the next subsection we evaluate Eq. \eqref{eq:interacting-fixed-momentum} on the fixed ghost direction supplied by the boundary BRST transformation.

Along the boundary path, the exact graph differential gives
\begin{align}
 g^{-1}\partial_\sigma g
 &=a_\sigma+\Gamma_\Phi(V_\sigma^\bullet),
 \nonumber\\
 K_\sigma^\bullet
 &=\mathcal T_\Phi(V_\sigma^\bullet),
 &
 K_\sigma^\circ
 &=\mathcal N_\Phi(V_\sigma^\bullet)
 =\mathcal D_\Phi(K_\sigma^\bullet).
 \label{eq:interacting-differentiated-graph}
\end{align}
The grade-zero connection is
$A_\sigma=a_\sigma+P_0\Gamma_\Phi(V_\sigma^\bullet)$ and generally has
both fixed and anti-fixed parts.  This is a kinematic consequence of
the graph condition, not an additional stationarity equation.

\subsection{Interacting boundary superfield equations}

Having imposed ordinary boundary stationarity, we now ask whether the
resulting boundary conditions also make the BRST flux vanish.  The
BRST calculation will use the fixed momentum condition
\eqref{eq:interacting-fixed-momentum} evaluated on the allowed ghost
direction $U^\bullet=\lambda_+^\bullet$.  This specialization is made
only after the equation has been derived for arbitrary $U^\bullet$; it
does not define a new variational equation.

The canonical momentum is the coefficient $L_\tau$ of the pulled-back
form $\star L$.  Its odd components satisfy
\begin{align}
 \mathcal P_1&=K_{1,z}+3K_{1,\bar z},
 &
 \mathcal P_3&=3K_{3,z}+K_{3,\bar z},
 \nonumber\\
 K_{1,\sigma}&=K_{1,z}-K_{1,\bar z},
 &
 K_{3,\sigma}&=K_{3,z}-K_{3,\bar z}.
 \label{eq:odd-momentum-and-tangent-current}
\end{align}
Consequently,
\begin{align}
 K_{3,z}
 =\frac14\left(\mathcal P_3+K_{3,\sigma}\right),
 \qquad
 K_{1,\bar z}
 =\frac14\left(\mathcal P_1-K_{1,\sigma}\right).
 \label{eq:odd-currents-from-momentum}
\end{align}
The unreduced BRST flux
\eqref{eq:brst-current-matching} can therefore be written as
\begin{align}
 4\mathcal B_{\rm PS}
 =\St\left[
 (\lambda+\bar\lambda)\mathcal P
 +(\lambda-\bar\lambda)K_\sigma
 \right].
 \label{eq:unreduced-interacting-BRST-flux}
\end{align}
The grade-two component vanishes automatically in the supertrace with
the odd ghosts.

Using Eq.~\eqref{eq:interacting-ghost-gluing}, we have
\begin{align}
 (\lambda+\bar\lambda)^\bullet&=\lambda_+^\bullet,
 &
 (\lambda+\bar\lambda)^\circ
 &=-\mathcal M^\circ
 \left(\lambda_+^\bullet,\lambda_-^\circ\right),
 \nonumber\\
 (\lambda-\bar\lambda)^\bullet
 &=\mathcal M^\bullet
 \left(\lambda_+^\bullet,\lambda_-^\circ\right),
 &
 (\lambda-\bar\lambda)^\circ&=-\lambda_-^\circ.
 \label{eq:ghost-combinations-from-M}
\end{align}
It follows that
\begin{align}
 4\mathcal B_{\rm PS}
 =\St\Big[
 \lambda_+^\bullet\mathcal P^\bullet
 -\mathcal M^\circ
  \left(\lambda_+^\bullet,\lambda_-^\circ\right)\mathcal P^\circ
 +\mathcal M^\bullet
  \left(\lambda_+^\bullet,\lambda_-^\circ\right)K_\sigma^\bullet
 -\lambda_-^\circ K_\sigma^\circ
 \Big].
 \label{eq:BRST-flux-before-surface-equations}
\end{align}

We now evaluate the full fixed momentum condition
\eqref{eq:interacting-fixed-momentum} at
$U^\bullet=\lambda_+^\bullet$.  Solving it for the first term in
Eq.~\eqref{eq:BRST-flux-before-surface-equations} gives
\begin{align*}
 \St\left(\lambda_+^\bullet\mathcal P^\bullet\right)
 ={}&
 -\St\left[
 \lambda_+^\bullet\mathbf F_{\Phi,\sigma}^\bullet
 +\mathcal D_\Phi(\lambda_+^\bullet)\mathcal P^\circ
 \right]
 \\
 &+\St\Bigg[
 \omega_-^\bullet
 \left(
 \nabla_{\lambda_+^\bullet}^\Phi\mathcal M^\bullet\right)
 \left(\lambda_+^\bullet,\lambda_-^\circ\right)
 \\
 &\hspace{14mm}
 +\omega_+^\circ
 \left(
 \nabla_{\lambda_+^\bullet}^\Phi\mathcal M^\circ\right)
 \left(\lambda_+^\bullet,\lambda_-^\circ\right)
 \Bigg].
\end{align*}
Equation~\eqref{eq:interacting-differentiated-graph} and the definition
of the supertrace adjoint give
\begin{align*}
 -\St\left(\lambda_-^\circ K_\sigma^\circ\right)
 &=-\St\left[
 \lambda_-^\circ\mathcal D_\Phi(K_\sigma^\bullet)\right]
 \\
 &=-\St\left[
 K_\sigma^\bullet
 \mathcal D_\Phi^\dagger(\lambda_-^\circ)\right].
\end{align*}
Finally, the force convention
\eqref{eq:embedded-curvature-force} gives
\begin{align}
 \St\left(\lambda_+^\bullet\mathbf F_{\Phi,\sigma}^\bullet\right)
 =\St\left(
 K_\sigma^\bullet\iota_{\lambda_+^\bullet}\mathbf F_\Phi\right).
 \label{eq:open-curvature-contraction}
\end{align}
The boundary momentum \(\mathcal P^\circ\) and the current \(K_\sigma^\bullet\) have even total Grassmann parity, so moving them to the left inside the supertrace introduces no additional graded sign.
Substitution of these three identities into
Eq.~\eqref{eq:BRST-flux-before-surface-equations} yields
\begin{align}
 4\mathcal B_{\rm PS}
 ={}&
 \St\Bigg\{
 K_\sigma^\bullet\left[
 \mathcal M^\bullet
  \left(\lambda_+^\bullet,\lambda_-^\circ\right)
 -\iota_{\lambda_+^\bullet}\mathbf F_\Phi
 -\mathcal D_\Phi^\dagger(\lambda_-^\circ)
 \right]
 \nonumber\\
 &\hspace{18mm}
 -\mathcal P^\circ\left[
 \mathcal M^\circ
  \left(\lambda_+^\bullet,\lambda_-^\circ\right)
 +\mathcal D_\Phi(\lambda_+^\bullet)
 \right]
 \Bigg\}
 \nonumber\\
 &+\St\Bigg[
 \omega_-^\bullet
 \left(
 \nabla_{\lambda_+^\bullet}^\Phi
 \mathcal M^\bullet\right)
 \left(\lambda_+^\bullet,\lambda_-^\circ\right)
 \nonumber\\
 &\hspace{14mm}
 +\omega_+^\circ
 \left(
 \nabla_{\lambda_+^\bullet}^\Phi
 \mathcal M^\circ\right)
 \left(\lambda_+^\bullet,\lambda_-^\circ\right)
 \Bigg].
 \label{eq:completed-interacting-BRST-flux}
\end{align}
The curvature contributes to the coefficient of the independent
tangential current in the embedded fixed frame.

The coefficients of $K_\sigma^\bullet$ and $\mathcal P^\circ$ must
vanish independently.  For every allowed pair
$(\lambda_+^\bullet,\lambda_-^\circ)$ inherited from the free boundary
pure-spinor data, the interacting boundary superfields therefore obey
\begin{align}
 \mathcal M_{\circ\bullet}(\lambda_+^\bullet)
 +\mathcal M_{\circ\circ}(\lambda_-^\circ)
 +\mathcal D_\Phi(\lambda_+^\bullet)
 &=0,
 \label{eq:final-embedding-equation}\\
 \mathcal M_{\bullet\bullet}(\lambda_+^\bullet)
 +\mathcal M_{\bullet\circ}(\lambda_-^\circ)
 -\iota_{\lambda_+^\bullet}\mathbf F_\Phi
 -\mathcal D_\Phi^\dagger(\lambda_-^\circ)
 &=0,
 \label{eq:final-curvature-equation}
\end{align}
The remaining antighost contribution gives a single equation on the antighost gauge quotient, rather than separate coefficient equations for $\omega_-^\bullet$ and $\omega_+^\circ$. The reconstruction \eqref{eq:boundary-ghost-reconstruction} must remain in $\mathfrak g_1\oplus\mathfrak g_3$, while the residual connection entering $\nabla^\Phi$ has grade zero. Consequently, the covariant variation of the reconstructed pair is grade-compatible, and its fixed and anti-fixed components are the two antighost coefficients in Eq. $\eqref{eq:completed-interacting-BRST-flux}$. Because the reconstructed pure-spinor constraints hold pointwise on the brane, this variation is tangent to the allowed pure-spinor cone. Moreover, $\omega_-^\bullet$ and $\omega_+^\circ$ are the corresponding $R$-parity components of one grade-one antighost class. On a smooth local patch, non-degeneracy of the quotient pairing therefore requires both components of the covariant variation to vanish:
\begin{subequations}
\label{eq:final-ghost-equation}
\begin{align}
 \left(
 \nabla_{\lambda_+^\bullet}^\Phi
 \mathcal M_{\bullet\bullet}\right)(\lambda_+^\bullet)
 +\left(
 \nabla_{\lambda_+^\bullet}^\Phi
 \mathcal M_{\bullet\circ}\right)(\lambda_-^\circ)
 &=0,
 \label{eq:final-ghost-fixed-equation}\\
 \left(
 \nabla_{\lambda_+^\bullet}^\Phi
 \mathcal M_{\circ\bullet}\right)(\lambda_+^\bullet)
 +\left(
 \nabla_{\lambda_+^\bullet}^\Phi
 \mathcal M_{\circ\circ}\right)(\lambda_-^\circ)
 &=0.
 \label{eq:final-ghost-antifixed-equation}
\end{align}
\end{subequations}
Equations \eqref{eq:final-ghost-fixed-equation} and \eqref{eq:final-ghost-antifixed-equation} are necessary consistency conditions for four couplings obeying the grade, cone, and antighost-gauge restrictions stated above. Their derivation does not by itself establish that the couplings extend from the allowed pure-spinor data to globally defined linear maps with all these properties. We have not assumed any functional dependence of $\mathcal M_{rs}$ on $\mathbf A^\bullet$ and $\Phi^\circ$. Instead, Eq. \eqref{eq:final-embedding-equation} determines the anti-fixed output combination from the embedding, while Eq. \eqref{eq:final-curvature-equation} determines the fixed output combination from the superconnection curvature and the adjoint embedding derivative. Both determined combinations are invariant under the Abelian open-string gauge symmetry because they involve the pulled-back curvature $\mathbf F_\Phi$, rather than $\mathbf A^\bullet$ itself. The last
two displayed equations are the $R$-parity components of the single
antighost-quotient equation; their compatibility is already contained
in the grade-preserving reconstruction
\eqref{eq:boundary-ghost-reconstruction}.

Equations~\eqref{eq:final-embedding-equation}--
\eqref{eq:final-ghost-equation} retain every power of $\Phi^\circ$
fixed by the local logarithmic coordinate.  They remain at the
Abelian level in $\mathbf A^\bullet$ and at quadratic order in the
boundary ghosts.  At
$\mathbf A^\bullet=\Phi^\circ=0$, the first two equations make both
output combinations vanish on the allowed pure-spinor cone and hence
recover the field-independent gluing used in the preceding sections.

\section{Conclusion and prospects}
\label{conclusion}

We have formulated a coordinate-independent algebraic framework for
zero-field Lorentzian boundary conditions in the
$AdS_5\times S^5$ pure spinor string. A compatible
involution $R$ pairs the odd $\mathbb Z_4$ grades,
splits the even target-space directions into Neumann and Dirichlet
sectors, and identifies a preserved symmetry superalgebra. The known
D1, D3, D5, and D7 probe geometries provide nontrivial targets for
checking this framework.
The matrix analysis supplies compatible representatives for the four unitary rows and the three orthosymplectic probe embeddings in the chosen basis, together with their real fixed algebras and grade-two dimensions. These seven representatives realize the
Lorentzian zero-flux D1, D3, D5, and D7 sectors summarized in
Section~\ref{sec:allowedDbranes}.  We also determined $G^R\cap H_0$ and used
the $\mathbb Z_4$ grades to distinguish the physical and stabilizer
Abelian directions in the D3 rows.  Grade exchange gives the required
pairing of the odd sectors, bracket preservation maps the pure-spinor
cone into itself, and
fixed/anti-fixed supertrace orthogonality proves the zero-field
boundary-BRST condition.  The classification of half-BPS supergravity
symmetry algebras in~\cite{DHoker:2008wvd} provides an important
cross-check of these world-sheet tests.

We have also derived the interacting boundary equations for four
$R$-parity-resolved ghost couplings that collectively preserve the
boundary pure-spinor cone and are compatible with the antighost gauge
quotient.  The local logarithmic coordinate $Y^\circ$, defined from the Cartan map, makes the first-order position term \eqref{eq:AdS-first-order-position} well defined.  Differentiating
the exponential graph gives the exact lift $\Gamma_\Phi$; its fixed
and anti-fixed coset blocks $\mathcal T_\Phi$ and $\mathcal N_\Phi$
define the nonlinear embedding operator
$\mathcal D_\Phi=\mathcal N_\Phi\mathcal T_\Phi^{-1}$.  Pulling the
curvature and covariant derivatives to this embedded frame keeps the
matter and ghost equations in a common set of variables.  The
quadratic ghost interaction
\eqref{eq:field-dependent-ghost-interaction} introduces the four
couplings without prescribing them as functionals of the matter
superfields.

With these restrictions,
Eqs.~\eqref{eq:final-embedding-equation}--
\eqref{eq:final-ghost-equation} follow from ordinary boundary
stationarity and the unreduced boundary BRST flux.  They are necessary
local consistency equations on the allowed boundary pure-spinor data
inherited from the zero-field gluing. The transverse embedding determines the anti-fixed output combination, whereas the pulled-back Abelian curvature and the adjoint embedding derivative determine the fixed combination. 
Equations \eqref{eq:final-ghost-fixed-equation} and \eqref{eq:final-ghost-antifixed-equation} are the two $R$-parity components of a single grade-preserving equation on the antighost quotient; they constrain the fixed-direction variations in a local antighost gauge slice.
Cone preservation and
compatibility with the antighost quotient remain additional
restrictions on the four couplings, and we have not attempted to solve
them for a general probe.  The zero-field limit recovers the fixed
gluing used in the probe classification.

The remaining problem is to solve these projected superspace equations for the allowed probe sectors and to determine their physical fluctuation content.  A component analysis for a specific D-brane probe, including the identification of its physical world-volume multiplet and nonlinear equations, will be presented in future work. Further extensions include a non-Abelian Chan--Paton completion and connections to holographic Wilson and defect operators.

\subsection*{Acknowledgements}
B.C.V. thanks Andrei Mikhailov and William Linch for discussions. Parts of this work were carried out with partial support from FONDECYT grants numbers 1151409 and
1250672.

\paragraph{Disclaimer.}
The author utilized GPT-5.6 Sol for structural organization, language refinement, assisted computations and critical review. The author takes full responsibility for the manuscript’s
content, including all ideas, mathematical derivations, and conclusions.

\appendix
\section{The \texorpdfstring{$\psu(2,2|4)$}{psu(2,2|4)} algebra}
\label{appendix:psu}

The $\psu(2,2|4)$ algebra is generated by $30$ bosonic and $32$
fermionic operators. Its bosonic subalgebra is $\su(2,2)\oplus\su(4)$.
We use two structural properties of this algebra. Its $\Zb_4$
decomposition $\frakg=\bigoplus_{j=0}^3\frakg_j$ underlies the
integrability of the sigma model
\cite{Bena:2003wd,Vallilo:2003nx}, while the vanishing of its dual
Coxeter number $h^\vee$ enters proofs of quantum conformal invariance
\cite{Vallilo:2002mh,Berkovits:2004xu}. The $\Zb_4$ operator $\Sigma$
acts on the decomposed algebra as
\begin{align}\label{z4eigenvalues}
  \Sigma(\frakg_j) = \im^j \frakg_j.
\end{align}
Equivalently, for $X=\sum_{j=0}^3X_j$ with
$X_j\in\frakg_j$,
\begin{align}
  \Sigma(X)=X_0+\im X_1-X_2-\im X_3,
  \qquad \Sigma^4={\bf 1},
\end{align}
and the grade projectors are
\begin{align}\label{eq:z4-projectors}
  P_j=\frac14\sum_{k=0}^3\im^{-jk}\Sigma^k.
\end{align}
These formulas fix the phase convention used in
$R\Sigma R^{-1}=\Sigma^{-1}$ throughout the main text.
We denote a generic basis by $\{\sT_A\}$; the range and meaning of the
index $A$ are specified in each realization below.
The Lie bracket and invariant supertrace pairing respect the $\Zb_4$
grading:
\begin{align}
  [\frakg_i,\frakg_j]&\subseteq \frakg_{i+j\, (\mathrm{mod}\,4)}, &
  \St(\frakg_i\frakg_j)&=0\quad\text{unless }i+j=0\ (\mathrm{mod}\,4).
\end{align}

We first describe $\mathfrak{psl}(4|4;\mathbb C)$ and then impose the reality condition defining $\mathfrak{psu}(2,2|4)$. We write a complex $(4|4)\times(4|4)$ supermatrix in block form as
\begin{align}
  M =\left(\begin{array}{cc} A & \Psi \\ \Theta & B\end{array}\right),
\end{align}
where $A$ and $B$ are bosonic matrices and $\Psi$ and $\Theta$ are
fermionic ones. The complex supermatrix belongs to $\mathfrak{sl}(4|4)$ if
it satisfies
\begin{align}
  \St(M)\equiv\Tr(A) -\Tr(B) =0.
\end{align}
The projective superalgebra $\mathfrak{psl}(4|4)$ is obtained by
quotienting by the central identity,
\begin{align}
  M \sim M + c\,{\bf 1}_8,\qquad c\in\mathbb C,
\end{align}
where ${\bf 1}_8$ is the $8\times 8$ identity matrix. The supertranspose is
\begin{align}
  M^\ST= \left(\begin{array}{cc} A^\T & \Theta^\T \\ -\Psi^\T & B^\T\end{array}\right).
\end{align}
This is the convention of Ref.~\cite{DHoker:2008wvd}; in particular,
$(M^\ST)^\ST$ changes the sign of both odd blocks.

To define the real form, introduce
\begin{align}
  I_{2,2}=\operatorname{diag}({\bf 1}_2,-{\bf 1}_2),\qquad
  L_{2,2|4}=\operatorname{diag}(I_{2,2},-\im {\bf 1}_4).
\end{align}
Following Eq.~(4.13) of Ref.~\cite{DHoker:2008wvd},
$\mathfrak{su}(2,2|4)$ consists of the matrices
$M\in\mathfrak{sl}(4|4;\mathbb C)$ satisfying
\begin{align}\label{eq:su224-reality}
  M=-L_{2,2|4}^{-1}(M^*)^\ST L_{2,2|4},
  \qquad\text{equivalently}\qquad
  (M^*)^\ST L_{2,2|4}+L_{2,2|4}M=0.
\end{align}
A scalar multiple of the identity satisfies this reality condition only when its coefficient is imaginary. Therefore
\begin{align}
  \psu(2,2|4)=\su(2,2|4)/(\im\mathbb R\,{\bf 1}_8),
\end{align}
while the complex projective quotient above remains by
$\mathbb C\,{\bf 1}_8$.

One useful realization makes the $\Zb_4$ symmetry and the subalgebra
$\frakg_0=\mathfrak{usp}(2,2)\oplus\mathfrak{usp}(4)\approx\so(1,4)\oplus\so(5)$ explicit.
Here $\spp(2)$ denotes the compact real form
$\mathfrak{usp}(4)\simeq\so(5)$.
The remaining generators transform in vector and spinor
representations of $\frakg_0$. In this realization, each subspace is
generated by
\begin{align}
  \frakg_0=\{ \sM_{ab},\sM_{a'b'}\},\quad\frakg_1=\{\sQ_\alpha\},\quad
  \frakg_2=\{\sP_a,\sP_{a'}\},\quad\frakg_3=\{\bar\sQ_\alpha\},
\end{align}
where $a=0\cdots 4$, $a'=5\cdots 9$ and $\alpha=1\cdots 16$. We take the generators in $\mathfrak g_0$ to be anti-Hermitian and all remaining generators to be Hermitian. The non-vanishing (anti-)commutators are
\begin{align}\label{algebra}
 [ \sM_{\underline{ab}} , \sM_{\underline{cd}} ]
 &= - \eta_{ \underline{a} [\underline{c} } \sM_{\underline{d} ]
   \underline{b} }
   + \eta_{ \underline{b} [\underline{c} } \sM_{\underline{d} ]
   \underline{a} } ,&
   [ \sM_{\underline{ab}} , \sP_{\underline{c}} ] &=
   \eta_{\underline{c} [ \underline{a}} \sP_{\underline{b}]} ,&&\\
 [ \sP_a , \sP_b ] &= - \sM_{ab},& [ \sP_{a'} , \sP_{b'} ] &= \sM_{a'b'} , \\
 [\sM_{\underline{ab}} , \sQ_\a ] &= \frac12
  (\g_{\underline{ab}})_\a{}^\b\sQ_\b,&
 [ \sM_{\underline{ab}} , \bar\sQ_\a ] &=\frac12
 (\g_{\underline{ab}})_\a{}^\b\bar\sQ_\b ,\\
 [ \sP_{\underline{a}} , \sQ_\a ] &= \frac{\im}{2}
 (\g_{\underline{a}}\bar\eta)_\a{}^{\g}\bar\sQ_\g,&
 [ \sP_{\underline{a}} , \bar\sQ_\a ] &= -\frac{\im}{2}
 (\eta\bar\g_{\underline{a}})_\a{}^{\g} \sQ_\g, \\
 \{ \sQ_\a , \sQ_\b \} &=  \g^{\underline{a}}_{\a\b}\sP_{\underline{a}},&
 \{ \bar\sQ_\a , \bar\sQ_\b \} &=  \g^{\underline{a}}_{\a\b}\sP_{\underline{a}}  ,\\
 \{ \sQ_\a , \bar\sQ_\b \} &= \frac{\im}{2} (\g^{ab}\eta)_{\a\b}
 \sM_{ab}- \frac{\im}{2} (\g^{a'b'}\eta)_{\a\b} \sM_{a'b'},
\end{align}
Here antisymmetrization has unit weight:
$X_{[\underline a}Y_{\underline b]}
=X_{\underline a}Y_{\underline b}
-X_{\underline b}Y_{\underline a}$.
We use $\underline{a}=\{a,a'\}$, while $(\gamma^{\underline
  a}_{\a\b},\bar\g_{\underline a}^{\a\b})$ are
ten-dimensional symmetric Majorana-Weyl gamma matrices and $\bar\eta
=\bar\g_0\g_1\bar\g_2\g_3\bar\g_4=\bar\eta^{\sT}$, $\eta
=\g_0\bar\g_1\g_2\bar\g_3\g_4=\eta^{\sT}$. Note that $\eta^{-1}=\bar\eta$.

Another useful description makes the $d=4$ conformal algebra $(\mathsf P_m,\mathsf M_{mn},\mathsf K_m,\mathsf D)$ explicit, with $m,n=0,\ldots,3$.
In this basis the $\Zb_4$ symmetry is no longer
manifest. The full algebra is obtained by adding the supercharges
together with their conformal counterparts and the $\su(4)$ generators
$(\sQ_\a^I,\bar\sQ_{\dot\a I},\sS_{\a I},\bar\sS_{\dot\a}^I,
\sR_I{}^J)$. Here the Greek indices are $SL(2,\mathbb{C})$ indices and
$I,J=1\cdots 4$ are $SU(4)$ indices. The $\su(4)$ generators are
traceless $\sR_I{}^I=0$. It is useful to use the four-dimensional
hermitian $\sigma$-matrices
$(\sigma^m_{\a\dot\a},\bar\sigma^{\dot\a\a}_m)$ and the invariant
tensors $(\epsilon_{\a\b},\epsilon_{\dot\a\dot\b})$ of
$SL(2,\mathbb{C})$ to define
\begin{align}
  \sP_{\a\dot\a}=\sigma^m_{\a\dot\a}\sP_m,\quad
  \sK_{\a\dot\a}=\sigma^m_{\a\dot\a}\sK_m,\quad
  \epsilon_{\dot\a\dot\b}\sM_{\a\b} -
  \epsilon_{\a\b}\sM_{\dot\a\dot\b} =
  -2\im\sigma^m_{\a\dot\a}\sigma^n_{\b\dot\b} \sM_{mn}.
\end{align}
We can use the tensor $\epsilon_{\a\b}$ and its inverse
$\epsilon^{\a\b}$ to lower and raise indices with the ordering
conventions $\sT^\a=\epsilon^{\a\b}\sT_\b$, $\sT_\a
=\epsilon_{\a\b}\sT^\b$. Generators with indices $\sT_\g$,
$\sT_{\dot\g}$, $\sT_L$ and $\sT^L$ will have the following
commutators with $\sM_{\a\b}$, $\sM_{\dot\a\dot\b}$ and $\sR_I{}^J$
\begin{align}
  [\sM_{\a\b},\sT_\g]&=\epsilon_{\g\a}\sT_\b +\epsilon_{\g\b}\sT_{\a},
  \quad [\sM_{\dot\a\dot\b},\sT_{\dot\g}]=
  \epsilon_{\dot\g\dot\a}\sT_{\dot\b}
  +\epsilon_{\dot\g\dot\b}\sT_{\dot\a}, \\
  [\sR_I{}^J, \sT_L ] &=\delta^J_L \sT_I -\frac14 \delta^J_I\sT_L,
  \quad  [\sR_I{}^J, \sT^L ] =-\delta^L_I \sT^J +\frac14 \delta^J_I\sT^L.
\end{align}
The commutator of $\sD$ with any other
generator $\sT$ gives the imaginary unit times the
canonical dimension of $\sT$. The canonical
dimensions of $(\sP,\sK,\sM,\sQ,\sS,\sR)$ are
$(1,-1,0,\frac12,-\frac12,0)$. The other non-vanishing
(anti-)commutators which are not implied by the rules described above are
\begin{align}
 \{\sQ^I_\alpha,\bar\sQ_{\dot\a J}\} = \delta^I_J\sP_{\a\dot\a},\quad
 \{\sS_{\a J},\bar\sS^I_{\dot\beta}\} = \delta^I_J \sK_{\a\dot\beta},\\
 \{\sQ^I_\a,\sS_{\beta J}\}\ =\delta^I_J(\sM_{\alpha\beta}+\frac{\im}{2}\epsilon_{\alpha\beta}\sD)
   -\epsilon_{\alpha\beta}\sR_J{}^I,\\
 {[\sP_{\alpha\dot\a},\sS_{\b I}]}= \epsilon_{\a\beta}\bar\sQ_{\dot\a I},\quad
{[\sK_{\a\dot\alpha}, \sQ_\b^I]}=\epsilon_{\a\b} \bar\sS_{\dot\a}^I ,\\
 {[\sP_{\alpha\dot\a},\sK_{\b\dot\b}]}\ =-\epsilon_{\dot\alpha\dot\b}\sM_{\a\beta}
  +\epsilon_{\a\beta}\sM_{\dot\alpha\dot\b}-
  \im\epsilon_{\a\beta}\epsilon_{\dot\alpha\dot\b}\sD.
\end{align}
The remaining commutators are implied by the reality conditions
\begin{align}
  &\left(\sR_I{}^J \right)^\dagger =\sR_J{}^I,\quad
  \sP_{\a\dot\a}^\dagger=\sP_{\a\dot\a},\quad
  \sK_{\a\dot\a}^\dagger=\sK_{\a\dot\a}, \quad \sD^\dagger=\sD,\\
  &\sM_{\a\b}^\dagger=\sM_{\dot\a\dot\b}, \quad
  \left( \sQ_\a^I\right)^\dagger =\bar\sQ_{\dot\a I},\quad
   \left( \sS_{\a I}\right)^\dagger =\bar\sS_{\dot\a}^I.
\end{align}

This description displays the subalgebras relevant to
four-dimensional physics, such as the Poincar\'e algebra and the
$\mathcal N=4$ super-Poincar\'e algebra, which is
generated by $(\sP,\sM,\sQ,\bar\sQ)$.

\section{The \texorpdfstring{$\osp(4|4;\mathbb R)$ and
  $\osp(4^*|4)$}{osp(4|4;R) and osp(4*|4)} algebras}
\label{appendix:osp}

We follow the conventions of Sections~4.2--4.6 and
Appendix~B.3--B.4 of Ref.~\cite{DHoker:2008wvd}.  Let
$\mathcal V=\mathcal V_{\bar 0}\oplus\mathcal V_{\bar 1}$ have complex
dimension $(m|2n)$, and write
\begin{align}
 M=\begin{pmatrix}A&B\\ C&D\end{pmatrix},\qquad
 M^\ST=
 \begin{pmatrix}A^\T&C^\T\\
 -B^\T&D^\T\end{pmatrix}.
 \label{eq:osp-supertranspose}
\end{align}
Thus $(M^\ST)^\ST$ agrees with $M$ on the even blocks and changes the
sign of the odd blocks.  With
\begin{align}
 J_{2n}=\begin{pmatrix}0&{\bf 1}_n\\-{\bf 1}_n&0\end{pmatrix},
 \qquad
 \mathcal K_{m|2n}=\operatorname{diag}({\bf 1}_m,J_{2n}),
 \label{eq:osp-bilinear-form}
\end{align}
the complex orthosymplectic algebra is
\begin{align}
 \osp(m|2n;\mathbb C)
 =\left\{M\in\mathfrak{gl}(m|2n;\mathbb C)\;\middle|\;
 M^\ST \mathcal K_{m|2n}+\mathcal K_{m|2n}M=0\right\}.
 \label{eq:complex-osp}
\end{align}
In blocks, this condition reads
\begin{align}
 A^\T+A=0,\qquad
 D^\T J_{2n}+J_{2n}D=0,\qquad
 B=-C^\T J_{2n}.
\end{align}
Consequently its even subalgebra is
$\so(m;\mathbb C)\oplus\mathfrak{sp}(2n;\mathbb C)$, while the odd
subspace transforms in the tensor product of their defining
representations.  The orthosymplectic condition also implies
$\St M=0$, so there is a natural embedding
$\osp(4|4;\mathbb C)\subset\mathfrak{sl}(4|4;\mathbb C)$.  Since the
central identity is not orthosymplectic, projection to
$\mathfrak{psl}(4|4;\mathbb C)$ remains injective.

The two real forms needed here are obtained from the semilinear
involutions in Eq.~(4.13) of Ref.~\cite{DHoker:2008wvd}.  In the same
basis as above they may be written
\begin{align}
 \osp(4|4;\mathbb R)
 &=\left\{M\in\osp(4|4;\mathbb C)\;\middle|\;M^*=M\right\},
 \label{eq:osp44-real}\\
 \osp(4^*|4)
 &=\left\{M\in\osp(4|4;\mathbb C)\;\middle|\;
 M=-\widetilde{\mathcal K}_{4|4}^{-1}(M^*)^\ST
 \widetilde{\mathcal K}_{4|4}\right\},
 \qquad
 \widetilde{\mathcal K}_{4|4}=\operatorname{diag}(J_4,{\bf 1}_4).
 \label{eq:osp4star4-real}
\end{align}
Equivalently, the first line consists of real supermatrices satisfying
Eq.~\eqref{eq:complex-osp}.  Their even subalgebras are
\begin{align}
 \osp(4|4;\mathbb R)_{\bar 0}
 &=\so(4)\oplus\mathfrak{sp}(4;\mathbb R)
 \simeq \so(3)\oplus\so(3)\oplus\so(2,3),
 \label{eq:osp44-even}\\
 \osp(4^*|4)_{\bar 0}
 &=\so(4^*)\oplus\mathfrak{sp}(4)
 \simeq \so(2,1)\oplus\so(3)\oplus\so(5).
 \label{eq:osp4star4-even}
\end{align}
Here $\mathfrak{sp}(4)$ denotes the compact algebra
$\mathfrak{usp}(4)$, as in Ref.~\cite{DHoker:2008wvd}.  Each real form
has sixteen real odd generators.  The star in $4^*$ labels the
non-compact real form $SO(4^*)$, often written $SO^*(4)$; it is not
complex conjugation.

The real-form inclusions relevant for the $AdS_5\times S^5$ background
are
\begin{align}
 \psu(2,2|4)\supset\osp(4^*|4),\qquad
 \psu(2,2|4)\supset\osp(4|4;\mathbb R).
 \label{eq:osp-psu-embeddings}
\end{align}
They are the two orthosymplectic entries in Eq.~(4.18) of
Ref.~\cite{DHoker:2008wvd}.  The first specializes
$\mathfrak{osp}(2m^*|2n)\subset\mathfrak{su}(m,m|2n)$; the second
specializes $\mathfrak{osp}(m|2n;\mathbb R)\subset
\mathfrak{su}(m|n,n)$, with the order of the two $(4|4)$ blocks
interchanged in the latter.

The changes of basis used below are adapted from Appendix~B.4 of
Ref.~\cite{DHoker:2008wvd}.  Define
\begin{align}
 S&=\frac{1}{\sqrt2}
 \begin{pmatrix}{\bf 1}_2&-\im{\bf 1}_2\\
                -\im{\bf 1}_2&{\bf 1}_2\end{pmatrix},
 & S^*&=S^{-1},\\
 S_1&=\operatorname{diag}({\bf 1}_4,S),
 &S_2&=\operatorname{diag}(S^{-1},{\bf 1}_4).
 \label{eq:osp-basis-change}
\end{align}
The inverse in our definition of $S_2$ is chosen to match the ambient
reality convention \eqref{eq:su224-reality}; Ref.~\cite{DHoker:2008wvd}
uses $S$ in the corresponding displayed matrix.
For $M\in\osp(4|4;\mathbb R)$,
$S_1^{-1}MS_1\in\su(4|2,2)$; exchanging the parity-block labels gives
the embedding in $\su(2,2|4)$.  For
$M\in\osp(4^*|4)$, $S_2^{-1}MS_2\in\su(2,2|4)$ directly.  These
inclusions are necessary algebraic facts.  The representatives in
Eq.~\eqref{eq:three-osp-C} position both real forms so that the
associated boundary map also reverses the sigma-model grading; this
relative position is not fixed by the abstract real-form inclusion.

For completeness, Tables~2 and~10 of Ref.~\cite{DHoker:2008wvd}
associate $\osp(4^*|4)$ with the $AdS_2$ D1 and
$AdS_2\times S^4$ D5 probes, and $\osp(4|4;\mathbb R)$ with the
$AdS_4\times S^2$ D5 probe.  For the Lorentzian representatives
retained in this paper, their homogeneous super-orbits and bosonic
stabilizers are
\begin{align}
 \mathcal W_{\mathrm{D1}}
 &=\frac{OSp(4^*|4)}{SO(1,1)\times SO(3)\times SO(5)},
 &(\mathcal W_{\mathrm{D1}})_{\mathrm{bos}}&=AdS_2,
 \label{eq:d1-osp-orbit}\\
 \mathcal W_{\mathrm{D5}(2,4)}
 &=\frac{OSp(4^*|4)}{SO(1,1)\times SO(3)\times SO(4)},
 &(\mathcal W_{\mathrm{D5}(2,4)})_{\mathrm{bos}}
 &=AdS_2\times S^4,
 \label{eq:d5-24-osp-orbit}\\
 \mathcal W_{\mathrm{D5}(4,2)}
 &=\frac{OSp(4|4;\mathbb R)}
 {SO(1,3)\times SO(2)\times SO(3)},
 &(\mathcal W_{\mathrm{D5}(4,2)})_{\mathrm{bos}}
 &=AdS_4\times S^2.
 \label{eq:d5-42-osp-orbit}
\end{align}
The identifications of their bosonic bodies use
\begin{align}
 AdS_2&=SO(2,1)/SO(1,1),&
 S^4&=SO(5)/SO(4),\\
 AdS_4&=SO(2,3)/SO(1,3),&
 S^2&=SO(3)/SO(2).
\end{align}
The remaining $SO(3)$ factors act on normal directions and therefore
belong entirely to the stabilizer.  This also shows why the D1 and the
$AdS_2\times S^4$ D5 may have the same preserved superalgebra without
defining the same brane orbit: their stabilizers inside
$OSp(4^*|4)$ are different.

\subsection{Matrix representatives for boundary involutions}
\label{appendix:matrix-involutions}

The matrix conventions above give a finite-dimensional description of
the inner and outer representatives used in
Section~\ref{sec:matrix-involutions}.  We define
\begin{align}
 \Pi=\operatorname{diag}({\bf 1}_4,-{\bf 1}_4),
 \qquad (M^\ST)^\ST=\Pi M\Pi,
\end{align}
and let square brackets denote the class of a matrix in
$\mathfrak{psl}(4|4)$.  Two useful families of complex-linear
automorphisms are
\begin{align}
 R^{\rm con}_{U}([M])&=[UMU^{-1}],\\
 R^{\rm st}_{C}([M])&=[-C^{-1}M^\ST C],
 \label{eq:matrix-R-families}
\end{align}
where $U$ and $C$ are invertible even matrices.  Both preserve the
central line and therefore descend to the projective algebra.

For the conjugation family,
\begin{align}
 \big(R^{\rm con}_{U}\big)^2=\operatorname{Ad}_{U^2},
\end{align}
so it is involutive on the projective algebra when $U^2$ is scalar. A
convenient starting family is
\begin{align}
 U_{p,q}
 =\operatorname{diag}(u_p,u_q),\qquad
 u_r=\operatorname{diag}({\bf 1}_r,-{\bf 1}_{4-r}).
 \label{eq:Upq}
\end{align}
The complex fixed algebra is the projective image of
\begin{align}
 \mathfrak{s}\big(
 \mathfrak{gl}(p|q)\oplus
 \mathfrak{gl}(4-p|4-q)\big).
\end{align}
Its intersection with the ambient real form produces the unitary
fixed algebras.  For the equal $2|2$ splitting, the two relevant real
forms give the D3 algebras listed in
Eq.~\eqref{eq:unitary-involution-table}; their Abelian directions are
assigned to the grade-zero and grade-two sectors in the main text.

For the supertranspose family one finds
\begin{align}
 \big(R^{\rm st}_{C}\big)^2
   =\operatorname{Ad}_{H_C},\qquad
 H_C=C^{-1}C^\ST\Pi.
 \label{eq:outer-square}
\end{align}
Consequently $R^{\rm st}_{C}$ is involutive on
$\mathfrak{psl}(4|4)$ whenever $H_C$ is scalar.  A useful sufficient
condition is
\begin{align}
 C^\ST=\epsilon C\Pi,\qquad \epsilon=\pm1.
 \label{eq:C-symmetry}
\end{align}
For $\epsilon=1$ the first diagonal block of $C$ is symmetric and the
second is antisymmetric; for $\epsilon=-1$ their roles are exchanged.

The canonical orthosymplectic representative is
\begin{align}
 J_4&=\begin{pmatrix}0&{\bf 1}_2\\-{\bf 1}_2&0\end{pmatrix},
 &\mathcal K_{4|4}&=\operatorname{diag}({\bf 1}_4,J_4),\\
 R_{\osp}([M])&=[-\mathcal K_{4|4}^{-1}M^\ST \mathcal K_{4|4}].
 \label{eq:Rosp-explicit}
\end{align}
In blocks this reads
\begin{align}
 R_{\osp}
 \begin{pmatrix}A&B\\C&D\end{pmatrix}
 =
 \begin{pmatrix}
  -A^\T&-C^\T J_4\\
  -J_4B^\T&J_4D^\T J_4
 \end{pmatrix}.
 \label{eq:Rosp-blocks}
\end{align}
The fixed-point equations are
\begin{align}
 A^\T+A&=0,&
 D^\T J_4+J_4D&=0,&
 C&=-J_4B^\T.
 \label{eq:osp-fixed-equations}
\end{align}
They are precisely
\begin{align}
 M^\ST \mathcal K_{4|4}+\mathcal K_{4|4}M=0,
\end{align}
and hence define $\osp(4|4;\mathbb C)$ in the conventions of
Ref.~\cite{DHoker:2008wvd}.  Under an even change of basis
$M'=V^{-1}MV$, the matrix defining the supertranspose-type involution
changes by congruence,
\begin{align}
 C'=V^\ST C V.
 \label{eq:C-congruence}
\end{align}
Equations~\eqref{eq:osp-basis-change} and~\eqref{eq:C-congruence}
therefore give explicit representatives in the corresponding unitary
bases.  In the $\osp(4|4;\mathbb R)$ case, the final parity-block
reordering also requires reordering the supertranspose and reality
matrices; $\mathcal K_{4|4}$ cannot simply be reused unchanged.

These complex fixed-point equations are only the first step.  Let the
ambient real form be the fixed set of
\begin{align}
 \phi_L(M)=-L_{\rm amb}^{-1}(M^*)^\ST L_{\rm amb}.
\end{align}
To verify the inner and outer representatives above as boundary
automorphisms of $\psu(2,2|4)$, we use the independent identities
\begin{align}\label{eq:R-matrix-checks}
 R^2&={\bf 1},&
 R\phi_L&=\phi_LR,&
 R\Sigma R^{-1}&=\Sigma^{-1},&
 \St(R(X)R(Y))&=\St(XY).
\end{align}
The real-form test distinguishes $\osp(4^*|4)$ from
$\osp(4|4;\mathbb R)$; the orthosymplectic fixed-point equation alone
does not.  The $\mathbb Z_4$ test is likewise independent and must be
performed in the precise sigma-model basis.

\paragraph{Construction and verification.}
In the common basis of Section~\ref{sec:matrix-involutions}, the four
inner representatives in Eq.~\eqref{eq:unitary-involution-table} and
the three outer forms in Eq.~\eqref{eq:three-osp-C} pass the
involution, real-form, and $\mathbb Z_4$ tests in
Eq.~\eqref{eq:R-matrix-checks}.  Diagonalizing their even action gives
\begin{align}
 \dim\frakg_2^\bullet=p+1
\end{align}
with the Lorentzian signatures and sphere dimensions displayed in the
two tables.  Their fixed real algebras have sixteen odd generators.
These checks establish the seven representatives used in the main
text.  Their isotropy intersections are displayed
in Eq.~\eqref{eq:probe-isotropy-table}, and the D3 Abelian directions
are resolved by Eq.~\eqref{eq:D3-abelian-grades}.

This calculation need not be done generator by generator.  Once a
basis $\{\sT_A\}$ and its dual are chosen, the full matrix of $R$ is
\begin{align}
 r_A{}^B=\kappa^{BC}\St\big(\sT_C R(\sT_A)\big),
 \qquad
 \kappa_{AB}=\St(\sT_A\sT_B).
 \label{eq:R-components}
\end{align}
The desired Neumann and Dirichlet signs can therefore be imposed as
algebraic constraints on $U$ or $C$, after which all fermionic images
can be displayed mechanically when explicit components are useful.
For existence, no row-by-row fermionic calculation is needed:
$R\Sigma R^{-1}=\Sigma^{-1}$ already gives
$R(\frakg_1)=\frakg_3$ and $R(\frakg_3)=\frakg_1$, while bracket
preservation gives
$\{R(\lambda),R(\lambda)\}=R(\{\lambda,\lambda\})=0$.
Together with the supertrace argument following
Eq.~\eqref{eq:tangent-normal-split}, this also establishes the
zero-field boundary-BRST condition.

{
\bibliographystyle{abe}
\bibliography{mybib}{}
}

\end{document}